\documentclass[twocolumn,showkeys,showpacs,superscriptaddress,preprintnumbers,amsmath,amssymb,aps]{revtex4-1}

\usepackage{graphicx}
\usepackage{dcolumn}
\usepackage{bm}

\begin{document}

\title{Decoherence of a quantum memory coupled to a collective spin bath}

\author{Richard Walters}
 \email{r.walters1@physics.ox.ac.uk}
 \affiliation{Clarendon Laboratory, University of Oxford, Oxford, OX1 3PU, UK}
\author{Stephen R. Clark}
 \affiliation{Centre for Quantum Technologies, National University of Singapore, 3 Science Drive 2, Singapore 117543}
 \affiliation{Clarendon Laboratory, University of Oxford, Oxford, OX1 3PU, UK}
\author{Dieter Jaksch}
 \affiliation{Clarendon Laboratory, University of Oxford, Oxford, OX1 3PU, UK}
 \affiliation{Centre for Quantum Technologies, National University of Singapore, 3 Science Drive 2, Singapore 117543}

\date{\today}

\begin{abstract}
We study the quantum dynamics of a single qubit coupled to a bath of interacting spins as a model for decoherence in solid state quantum memories. The spin bath is described by the Lipkin-Meshkov-Glick model and the bath spins are subjected to a transverse magnetic field. We investigate the qubit interacting via either an Ising- or an XY-type coupling term to subsets of bath spins of differing size. The large degree of symmetry of the bath allows us to find parameter regimes where the initial qubit state is revived at well defined times after the qubit preparation. These times may become independent of the bath size for large baths and thus enable faithful qubit storage even in the presence of strong coupling to a bath. We analyze a large range of parameters and identify those which are best suited for quantum memories. In general we find that a small number of links between qubit and bath spins leads to less decoherence and that systems with Ising coupling between qubit and bath spins are preferable.
\end{abstract}



\maketitle

\section{Introduction}
\label{Introduction}

The transfer and storage of quantum information between different
physical systems is of crucial importance to quantum communication
and distributed quantum computing~\cite{NielsenChuang}.
Experimentally quantum communication is one of the most advanced
areas of quantum technology with promising commercial applications
being a realistic possibility in the near
future~\cite{Gisin-quantcom-02}. However, for its full potential to
be realized over relevant distances, limitations caused by noisy
transmission lines, e.g. the exponential scaling of photon losses in
an optical fiber with its length, need to be overcome.

Several promising proposals to solve this problem exist. The most
well studied is the use of quantum repeaters which segment the
transmission line into shorter
pieces~\cite{Briegel-repeater-98,Duan-repeater-01}. By applying
sophisticated entanglement purification schemes and entanglement
swapping at these repeater units, high-fidelity entangled pairs can
be established over much larger distances than direct transmission
could feasibly permit. A recent alternative method has been proposed
that is based on the idea of entanglement percolation in a quantum
network. Given nodes that are initially connected by partially
entangled pure states~\cite{Acin-percolation-07} or mixed
states~\cite{Broadfoot-09}, schemes based on classical bond
percolation have been shown to enable the creation of maximally
entangled singlet states between arbitrary points in the network
with a probability independent of the distance between
them~\cite{Perseguers-percolation-08}. Both quantum repeater and
percolation schemes require that quantum information is stored
locally at the repeaters or nodes of the network and crucially their
operation degrades in the presence of decoherence in these memories.

In this work we investigate the quantum dynamics of a single qubit
quantum memory coupled to an environment of interacting spin-1/2
particles~\cite{RepProgPhys.63.699}. Such a model of decoherence has
relevance for solid-state quantum memories involving the nuclear
spin~\cite{PhysRevLett.91.246802, Morton-solidstate-08}, electron
spin in a semiconducting quantum dot \cite{PhysRevLett.83.4204} and
nitrogen vacancy centers in diamond \cite{L.Childress10132006}.
Previous studies examining the decoherence induced by spin baths
have mostly considered the so-called `central spin model' in which
the qubit is coupled isotropically to all spins in the bath. Early
work has analyzed the decoherence due to independent bath
spins~\cite{PhysRevD.26.1862, PhysRevA.66.052317,
PhysRevB.70.045323, PhysRevA.72.052113}, whilst most studies since
have investigated 1D models with nearest-neighbor couplings
\cite{PhysRevLett.96.140604, PhysRevA.75.032337, JPhysA.40.2455,
PhysRevB.75.094434, PhysRevB.77.205419, PhysRevA.75.012102}. The
effect of an infinite-range interaction amongst bath spins has also
been studied within the central spin model \cite{JPhysA.36.12305,
PhysRevA.76.012104, PhysRevB.76.174306, PhysRevA.78.060102}. However, as pointed out by
Rossini \textit{et al.} in Ref. \cite{PhysRevA.75.032333}, the
assumption of a central spin may not be valid for many physical
systems. Thus, in their work \cite{PhysRevA.75.032333} the authors
depart from the central spin model and consider just a few links
between the qubit and a 1D spin chain. Here we consider the spin
bath as being described by the Lipkin-Meshkov-Glick (LMG) model (see
Eq. \ref{H_LMG}), which possesses infinite-range interactions, and
similarly go beyond the central spin model by examining the
decoherence of a single qubit as its exposure to the spin bath is
varied from coupling to just one bath spin through to interacting
with all bath spins.

The LMG spin bath represents an ideal benchmark for investigating
the effects of the varying qubit interactions with a non-trivial
spin bath. Firstly, it possesses permutational symmetry, which can be
exploited to exactly solve its dynamics numerically for very large
number of spins ($\sim 100$-$1000$). This in itself is a rare
property for many-body quantum systems. The large degree of symmetry
in the bath also allows us to identify parameter regimes where
highly non-Markovian~\cite{PhysRevB.70.045323} coherence properties
are displayed within the quantum memory. We quantify this by its
rephasing time and periodic entanglement breaking, which are found
to be independent of the bath size. Secondly, the LMG model has also
attracted much attention \cite{PhysRevA.69.022107,
PhysRevA.69.054101, PhysRevA.71.064101, PhysRevA.77.052105,
PhysRevLett.101.025701} for exhibiting either first- or second-order
equilibrium quantum phase transitions dependent on the type of
intra-bath coupling. As is now well known the entanglement
properties of the ground state are strongly affected at criticality
\cite{Nature.416.608, PhysRevA.66.032110, Rev.Mod.Phys.80.517,
PhysRevLett.90.227902} and this in turn can have a profound effect
on the induced decoherence of a coupled
qubit~\cite{PhysRevLett.96.140604}. We explore the effects of
criticality and in a similar fashion to H. T. Quan \textit{et al.}
in Ref. \cite{PhysRevLett.96.140604} our findings point to a
possible use of a coupled quantum memory as an apparatus for the
dynamical measurement of a QPT through its coherence properties.

The structure of this paper is as follows. In Sec. \ref{model} we
introduce the model and the bath Hamiltonian, and describe the
ground state properties of the bath under various parameter regimes.
We also introduce the dephasing Ising interaction and dissipative XY
interaction terms and discuss the measures used to quantify
decoherence in our system. The results are presented in Sec.
\ref{Ising} and \ref{LMG} for dephasing and dissipative interactions
respectively, with conclusions drawn in Sec. \ref{conclusion}.

\section{The Model}
\label{model}

\begin{figure}
\includegraphics[width=6cm]{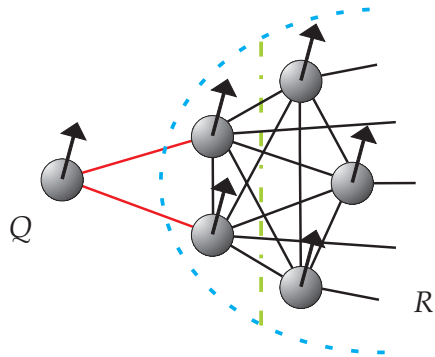}
\caption{(Color on line) A schematic of the model under investigation. A single
spin-1/2 is coupled via $k$ links (red lines) to a completely
connected spin bath $R$. Note that the lengths of the lines do not
correspond to the strength of the interactions. The green
dash-dotted line indicates a particular bath partition.}
\label{schematic}
\end{figure}

The model under consideration is that of a single spin-1/2 (qubit)
$Q$ interacting with a spin bath $R$. The purpose of the
investigation is two-fold: the primary focus is to study how the
interaction with the bath affects the evolution of $Q$, however in
doing so we also use $Q$ as a probe to study the bath dynamics
across criticality. The Hamiltonian of the global system, depicted
schematically in Fig. \ref{schematic} is of the standard form
\begin{equation}
H_T = H_Q + H_R + H_I \textrm{ ,}
\end{equation}
where $H_Q$ and $H_R$ are the free Hamiltonians of the qubit and the
bath respectively, and $H_I$ is the interaction term. For the qubit
we assume
\begin{equation}
H_Q = -\omega\varsigma^z \textrm{ ,}
\end{equation}
where $\varsigma^z$ is the Pauli $z$-operator and $\omega$ is the
energy difference between the ground and excited states; we work in
units with $\hbar=1$. In the following subsections, we will
introduce the bath Hamiltonian $H_R=H_{LMG}$ and the two different
interaction terms under consideration.  Throughout this work, the
subsystems $Q$ and $R$ will be initiated at time $t=0$ in a product
state, with the bath in its ground state.

\subsection{The Lipkin-Meshkov-Glick bath}
\label{LMG_bath}

The bath is described by the Lipkin-Meshkov-Glick model, which was
originally proposed in nuclear physics to describe shape phase
transitions in nuclei \cite{NuclPhys.62.188}. We note that it has
since found relevance in describing the Josephson effect in two-mode
Bose-Einstein condensates \cite{PhysRevA.57.1208} and a dissipative
implementation of it has also recently been proposed in optical
cavity QED \cite{PhysRevLett.100.040403}. The model involves a
system of $N$ spins-$1/2$ mutually interacting in the $x$-$y$ plane
and embedded in a transverse magnetic field. The Hamiltonian is
given by
\begin{eqnarray}
H_{LMG} & = & -\frac{\lambda}{N}\sum_{i<j}^N{(\sigma^x_i\sigma^x_j+\gamma\sigma^y_i\sigma^y_j)}-h\sum_{i=1}^N{\sigma^z_i} \nonumber \\
& = & -\frac{2\lambda}{N}(S^2_x+\gamma
S^2_y)-2hS_z+\frac{\lambda}{2}(1+\gamma) \textrm{ ,} \label{H_LMG}
\end{eqnarray}
where $S_\alpha=\sum_{i=1}^N{\sigma^\alpha_i/2}$ is the collective
spin operator for the direction $\alpha=x,y,z$; and the
$\sigma^\alpha$ are the Pauli operators for spins in the bath. The
quantities $\lambda$, $h$, and $\gamma$ characterize the interaction
strength between bath spins, the magnetic field strength, and the
anisotropy in the $x$-$y$ plane, respectively. The above Hamiltonian
describes an infinite-ranged XY model, with the prefactor $1/N$ on
the first term necessary to ensure a finite free energy per spin in
the thermodynamical limit. The cases $\gamma=1$ and $\gamma=0$
correspond to infinite-ranged XX and transverse Ising models
respectively. In the following, we consider only ferromagnetic
interactions $\lambda>0$ within the bath for anisotropy values
$0\leq\gamma\leq1$.

The mutual coupling of all spins described by $H_{LMG}$ results in
the energy eigenstates of the free bath being invariant under
particle exchange. Furthermore, $H_{LMG}$ commutes with the total
spin operator $\mathbf{S}^2=S_x^2+S_y^2+S_z^2$, and so the energy
eigenstates are divided into symmetric subspaces $\mathbb{S}_S$
defined by the spin quantum number $S$ of $\mathbf{S}^2$ and each of
dimensionality $2S+1$. This symmetry vastly reduces the complexity
of studies of the ground state (GS) of the bath. For ferromagnetic
intra-bath couplings the GS lies in the maximal spin subspace
$\mathbb{S}_{N/2}$, and so its calculation is an eigenvalue problem
scaling as just $N+1$.

For isotropic intra-bath coupling ($\gamma=1$) $H_{LMG}$ also
commutes with $S_z$ and is thus diagonal in the symmetric eigenbasis
spanned by Dicke states $|S,M\rangle$ \cite{PhysRev.93.99} ($M$ is
the quantum number for $S_z$). Dicke states are degenerate except
for those belonging to the maximal spin subspace $\mathbb{S}_{N/2}$.
For an ensemble of $N$ spins, Dicke states in the subspace
$\mathbb{S}_{N/2}$ are entangled states except for the two
spin-polarized states $|N/2,N/2\rangle$ and $|N/2,-N/2\rangle$.

\subsubsection{Ground state properties}

\begin{figure}
\includegraphics[width=4.1cm]{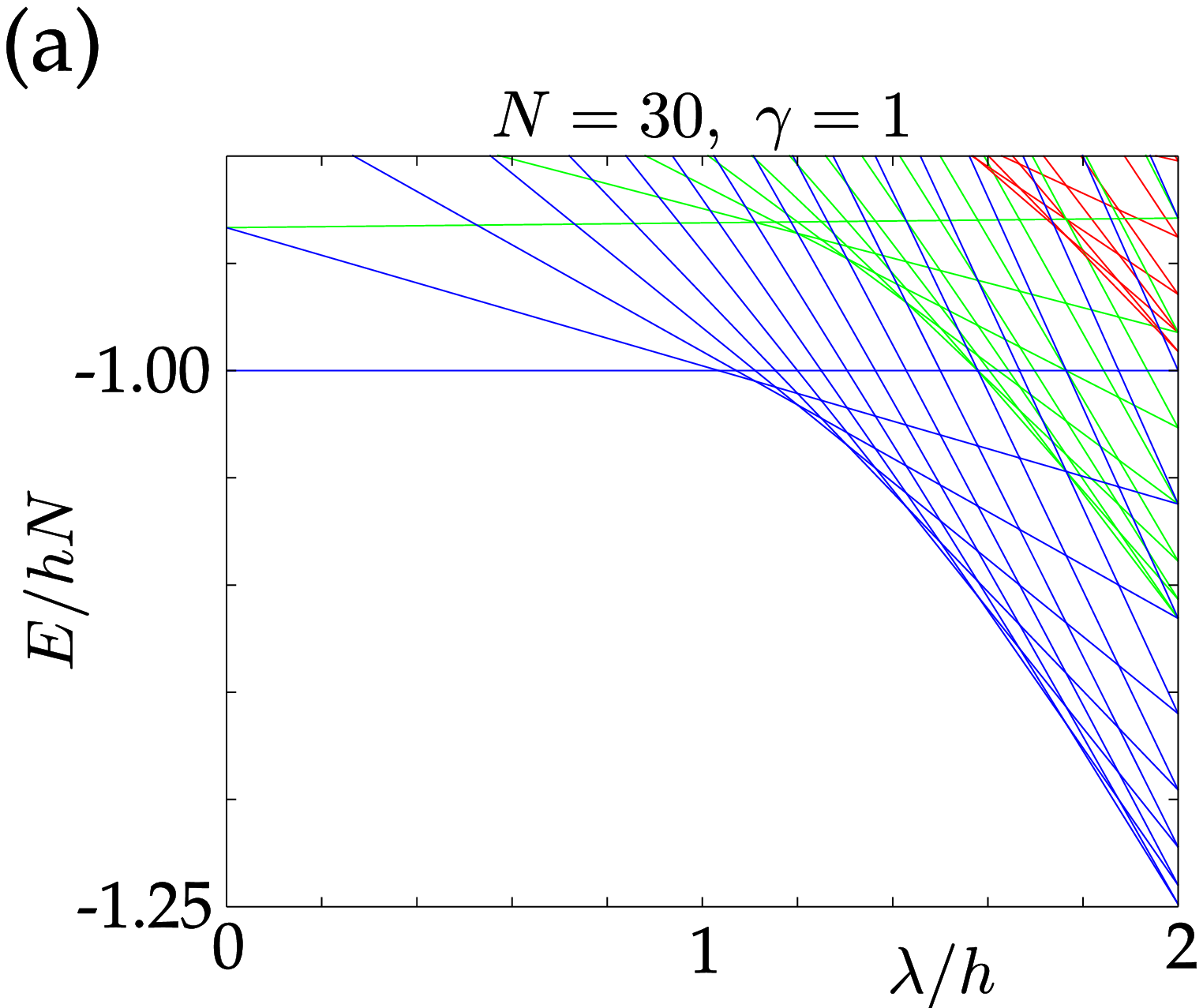}\hspace{0.25cm}
\includegraphics[width=4.1cm]{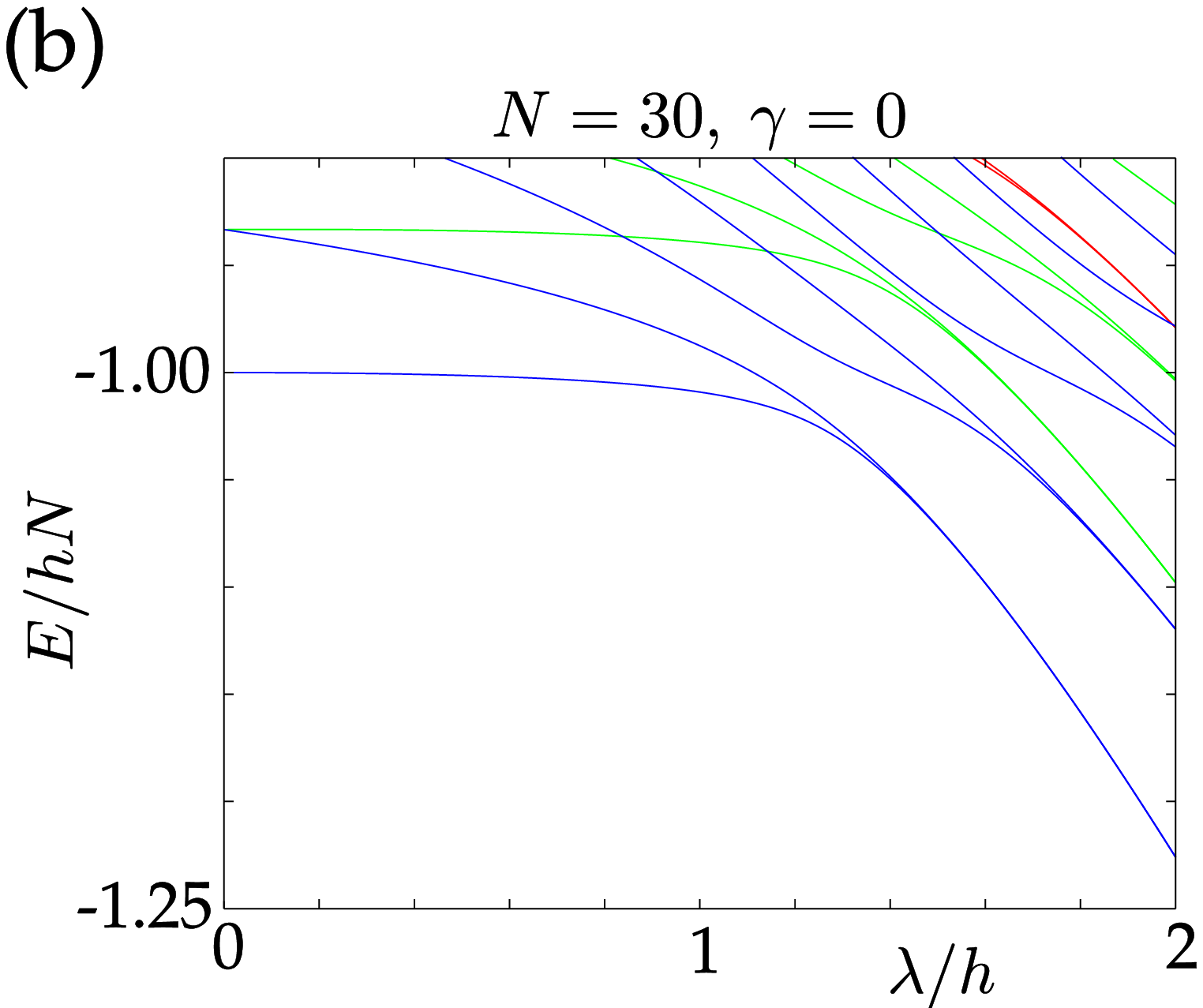}\vspace{0.3cm}
\includegraphics[width=4.1cm]{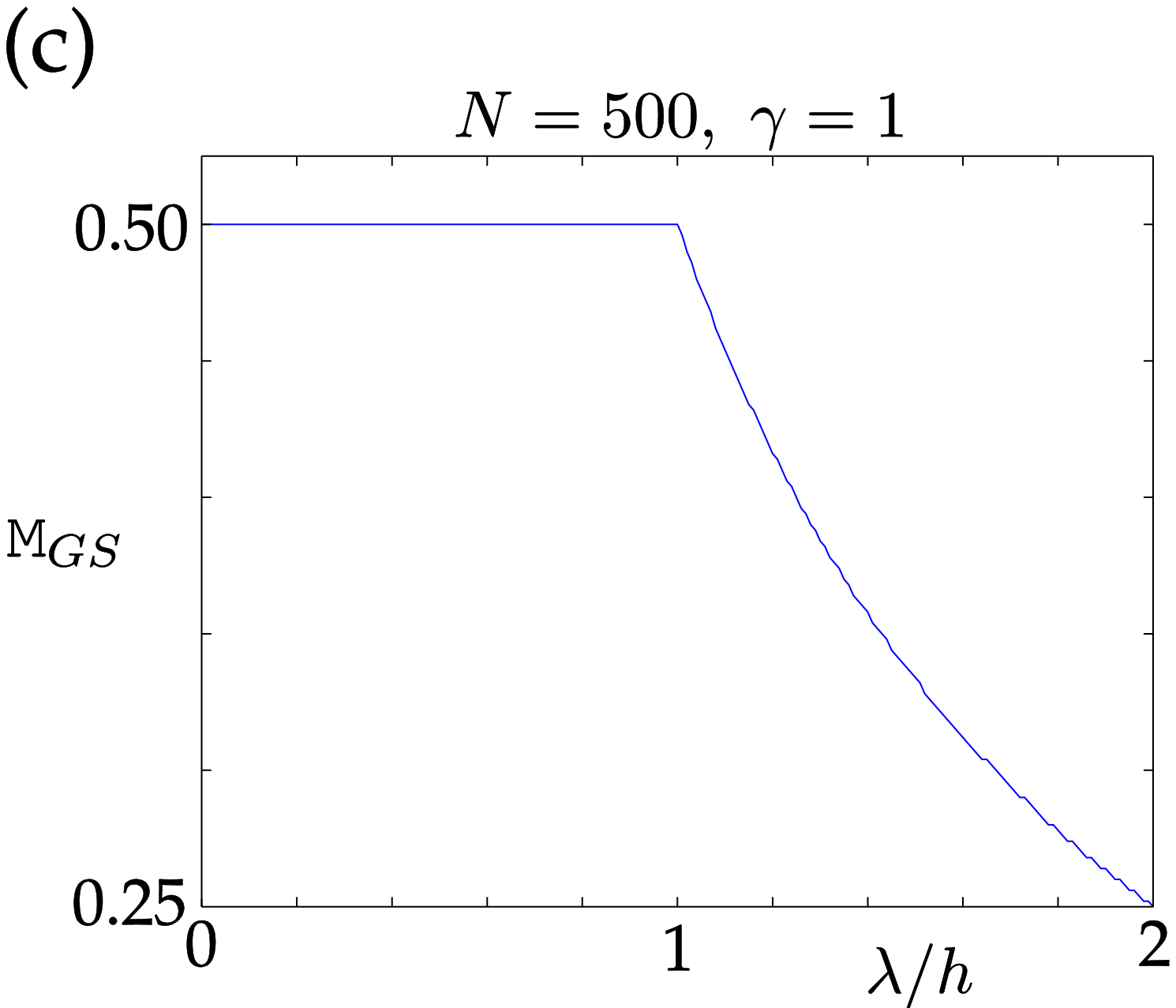}\hspace{0.25cm}
\includegraphics[width=4.1cm]{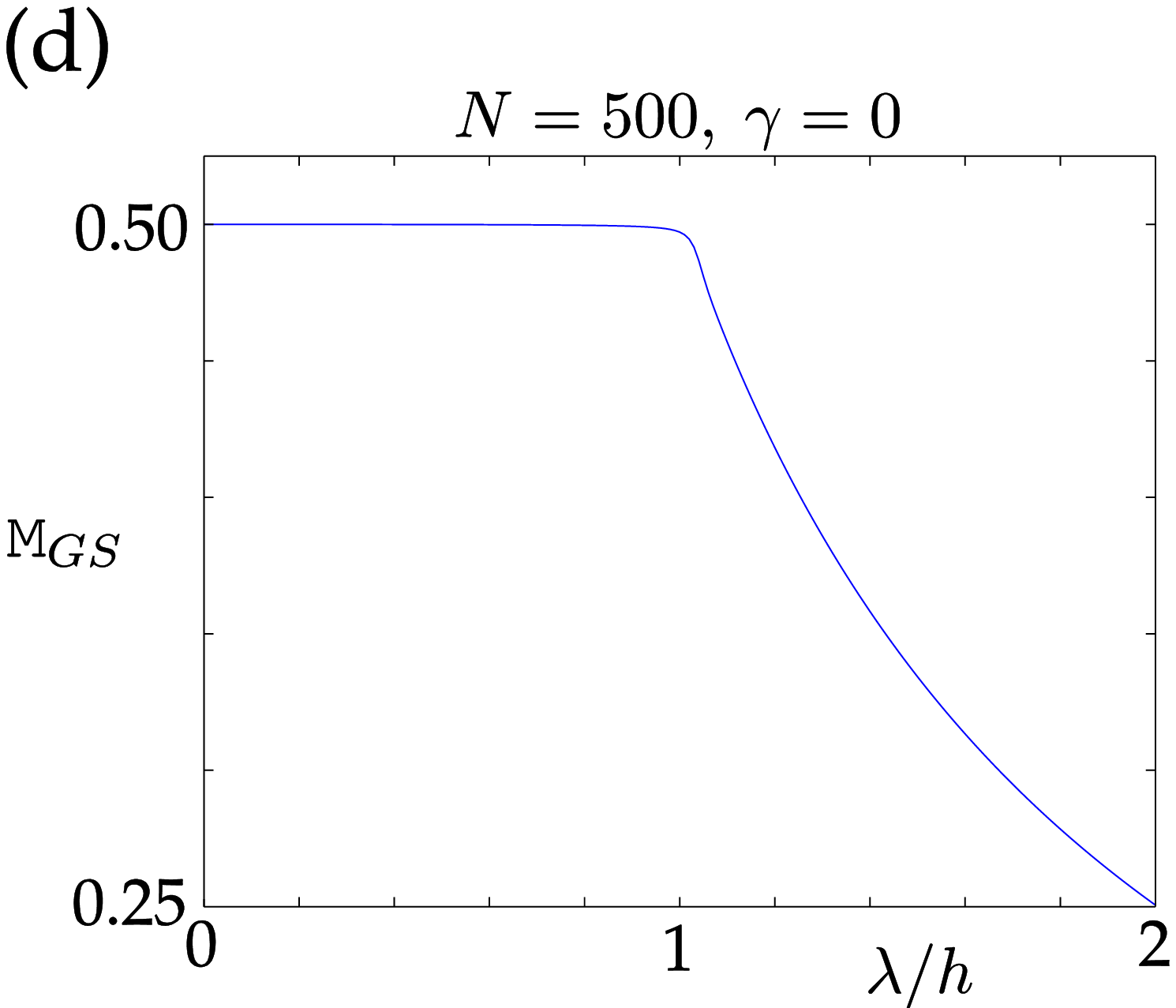}
\caption{(Color on line) Upper: the energy spectrum for isotropic (a) and
anisotropic (b) ferromagnetically coupled baths of $N=30$ spins.
Blue lines indicate levels in the subspace $\mathbb{S}_{N/2}$, green
lines the subspace $\mathbb{S}_{(N/2)-1}$, and red lines the
subspace $\mathbb{S}_{(N/2)-2}$. Lower: GS magnetization (per spin)
in the $z$-direction for isotropic (c) and anisotropic (d)
ferromagnetically coupled baths of $N=500$ spins.} \label{ferro}
\end{figure}

The ground state of the ferromagnetic bath lies in the maximal spin
subspace $\mathbb{S}_{N/2}$ for all $\gamma$ and $\lambda>0$. A
mean-field approach \cite{PhysRevLett.49.478, PhysRevB.28.3955}
predicts a second-order quantum phase transition in the
thermodynamical limit at the critical point $\lambda=h$ for all
$\gamma\geq 0$.

For a finite, isotropic bath there is a crossing of the lowest
energy levels at $\lambda_c =hN/(N-1)$. This is shown in Fig.
\ref{ferro}(a), where we have plotted the energies $E=\langle
H_{LMG}\rangle$ of the lowest levels for $\gamma=1$. The GS is the
separable, spin-polarized state $|N/2,N/2\rangle$ for
$0\leq\lambda\leq\lambda_c$ (normal phase), and an entangled state
$|N/2,\lfloor hN/2\lambda\rceil\rangle$ for $\lambda >\lambda_c$
(broken phase), where the function $\lfloor x\rceil$ rounds $x$ to
the nearest integer. For both phases the GS is non-degenerate for
any bath size. Note that the critical point depends on $h$, thus the
bath can be directed to either phase by setting the magnetic field
strength correspondingly greater or less than the intra-bath
coupling strength.

The GS of an anisotropic bath is a superposition over all Dicke
states in the maximal spin subspace $\mathbb{S}_{N/2}$, but tends to
the spin-polarized state $|N/2,N/2\rangle$ in the limit
$\lambda\rightarrow 0$. The GS is non-degenerate for the normal
phase, but becomes doubly degenerate in the thermodynamical limit
for the broken phase. This is demonstrated in Fig. \ref{ferro}(b),
which shows that the energies of the lowest levels are almost
identical for an anisotropic bath of just $N=30$ spins. In the
thermodynamical limit, the GS magnetization (per spin) in the
$z$-direction is given by
\begin{eqnarray}
\texttt{M}_{GS}=\frac{1}{N}\langle S_z\rangle = \left\{
\begin{array}{ccc}
1/2 & \textrm{for} & 0\leq\lambda\leq h \\
h/2\lambda & \textrm{for} & \lambda \geq h
\end{array}\right. \textrm{,}
\end{eqnarray}
for all $\gamma$. See Figs. \ref{ferro}(c) and \ref{ferro}(d) for the
magnetization of a large but finite bath with $\gamma=1$ and
$\gamma=0$ respectively.

\subsubsection{Analytical calculations with a partitioned LMG bath}
\label{partition}

The high symmetry of the LMG bath allows us to easily compute exact
solutions of the system dynamics for a large number of spins, since
the GS can be expressed in a reduced subspace $\mathbb{S}_S$ scaling
at most as $N+1$. However, this representation of the GS is limited
in that we can only consider operators that are symmetric across the
entire bath ensemble. Fortunately, the LMG model provides some
flexibility since we can make partitions of the bath, as shown in
Fig. \ref{schematic}, without significantly increasing the scaling
with system size, and thus can also consider non-symmetric
operators. As an example, consider an operator that acts on, and is
symmetric across, a total of $k$ bath spins. We make a bipartite
split of the bath into subsystems $A$ and $B$ with $k$ and $N-k$
spins respectively, and transform the original Dicke states as
\begin{eqnarray}
|(s_A,s_B)S,M\rangle & = & \sum_{m_A, m_B}{|s_A,m_A\rangle\otimes |s_B,m_B\rangle} \nonumber \\
& & \times\ \langle s_A,m_A|\otimes \langle s_B,m_B|S,M\rangle
\textrm{ ,} \label{ang}
\end{eqnarray}
where $|s_A,m_A\rangle$ and $|s_B,m_B\rangle$ are Dicke states for
subsystems $A$ and $B$; $s_A\leq k/2$; and $s_B\leq (N-k)/2$. We
denote the tensor product space by
$\mathbb{S}^k_{s_A}\otimes\mathbb{S}^{N-k}_{s_B}$, where the
superscripts indicate the corresponding number of spins. On the LHS
of Eq. (\ref{ang}) we specify $s_A$ and $s_B$ because, in general,
there are a range of allowed values these quantum numbers can take
when coupling to form $|S,M\rangle$. However, for the case where $S$
is maximal, which is our only concern, we have $s_A=k/2$,
$s_B=(N-k)/2$ and $s_A+s_B=S$, reducing the Clebsch-Gordan
coefficients in Eq. (\ref{ang}) from the general form in Ref.
\cite{JournCompPhys.122.343} to
\begin{widetext}
\begin{equation}
\langle k/2,m_A|\otimes\langle (N-k)/2,m_B|N/2,M\rangle =
\delta_{m_A+m_B,M}\sqrt{\frac{F_{m_A+k/2}(k)F_{m_B+(N-k)/2}(N-k)}{F_{M+N/2}(N)}}
\textrm{ ,} \label{CG1}
\end{equation}
where $F_b(a)=a!/[b!(a-b)!]$ are binomial coefficients. From Eqs.
(\ref{ang}) and (\ref{CG1}) we can construct a transformation
operator that maps the subspace $\mathbb{S}_{N/2}$ to the tensor
product space $\mathbb{S}^k_{k/2}\otimes\mathbb{S}^{N-k}_{(N-k)/2}$,
given by
\begin{multline}
T_{N/2\rightarrow k/2,(N-k)/2} = \sum^{N/2}_{M=-N/2}{\sum^{k/2}_{m_A=-k/2}{\sum^{(N-k)/2}_{m_B=-(N-k)/2}{\delta_{m_A+m_B,M}}}}\sqrt{\frac{F_{m_A+k/2}(k)F_{m_B+(N-k)/2}(N-k)}{F_{M+N/2}(N)}} \\
\times |k/2,m_A\rangle\otimes |(N-k)/2,m_B\rangle\langle N/2,M|
\textrm{ .} \label{T-op}
\end{multline}
\end{widetext}
We apply this transformation operator to the representations in
$\mathbb{S}_{N/2}$ of both the GS and $H_{LMG}$, allowing us to
consider operators that are symmetric across just $k$ spins without
needing to revert to the full Hilbert space $\mathbb{C}_2^{\otimes
N}$.

\subsection{Ising interaction}

The first interaction term we consider is of the Ising type
\begin{equation}
H_I =
\frac{\epsilon}{\sqrt{k}}\varsigma^z\otimes\sum_{i=1}^{k}{\sigma_i^z}
\textrm{ ,} \label{Ising_int}
\end{equation}
where $\epsilon$ is the interaction strength. This commutes with the
free qubit Hamiltonian (i.e. $[H_I,H_Q]=0$) and thus there will be
no energy exchange between the qubit and the bath; the evolution
will affect the off-diagonal terms of the state of $Q$ only. The
term above describes a situation involving a total of $k$ out of $N$
bath spins interacting with the qubit. By using the transformation
operator in Eq. (\ref{T-op}) we can vary the value of $k$ to study
any scenario from a single coupled bath spin ($k=1$) to a completely
coupled bath (i.e. the central spin model, $k=N$), allowing us to
evaluate how the evolution of the qubit changes as it is exposed to
varying numbers of bath spins. We have introduced the factor
$1/\sqrt{k}$ so as not to artificially increase the interaction
strength as we increase the number of links, effectively normalizing
the interaction term.

\subsection{LMG interaction}

With the second interaction we allow energy exchange between the
qubit and the bath and thus introduce dissipation into the subsystem
$Q$. A natural way of doing this is to consider the qubit
interacting with the bath spins in the same way the bath spins do
with each other, i.e.
\begin{equation}
H_I =
\frac{\epsilon}{\sqrt{k}}\sum_{i=1}^{k}{\left(\varsigma^x\otimes\sigma_i^x+\gamma\varsigma^y\otimes\sigma_i^y\right)}
\textrm{ .} \label{LMG_int}
\end{equation}
The anisotropy parameter $\gamma$ used here is the same as that in
the LMG Hamiltonian in Eq. (\ref{H_LMG}). Once again, we vary $k$
from $1$ to $N$ in order to consider different levels of exposure of
the qubit to the bath.

\subsection{Characterizing decoherence}

The main aim of this work is to characterize how the coupling to an
LMG spin bath affects the decoherent evolution of the qubit $Q$. In
the following subsections we introduce two measures that allow us to
do this; namely, the average purity, and whether the induced
decoherence is `entanglement breaking'. Their calculation requires
the use of an ancilla qubit and exploits the Jamiolkowski
isomorphism \cite{Jamiolkowski.3.275}, which is detailed in appendix
\ref{QOFandJI}. Importantly, all the information pertaining to the
decoherent evolution of the single qubit is mapped via the
Jamiolkowski isomorphism to a two-qubit state $\rho^\Lambda$ for $Q$
and the ancilla. From $\rho^\Lambda$ we can ascertain if and when
the evolution is entanglement breaking, and also deduce the Kraus
operators for the evolution to obtain the average purity.

\subsubsection{Average purity}

We use the purity of the reduced quantum state $\rho(t)$ of the
qubit, defined as $P[\rho]\equiv\textrm{Tr}(\rho^2)$, as a measure
of its decoherent evolution in the presence of the bath. The qubit
state at time $t$ will be pure iff $P(t)=1$, and mixed for $P(t)<1$.
We initialize the qubit in a pure state at $t=0$ so that any mixing
of the state, i.e. decoherence, can be observed in its subsequent
evolution described by the unital map $\varepsilon\{\rho(0)\}$. In
general, the purity depends on the exact details of the state at
$t=0$ and thus we average over all possible initial pure states
$|\psi\rangle$ to obtain the average purity. This can be achieved
once the Kraus operators $A_i$ for the evolution are known via
\cite{PhysRevA.70.012315}
\begin{eqnarray}
\overline{P} & = & \overline{\textrm{Tr}[\varepsilon\{|\psi\rangle\langle\psi |\}^2]}^\psi \nonumber \\
& = &
\frac{1}{d(d+1)}\left(\sum_{ij=1}^{d^2}\left|\textrm{Tr}[A_i^\dagger
A_j]\right|^2 + d\right) \textrm{ ,} \label{average_purity}
\end{eqnarray}
where $d=2$ is the dimension of the quantum subsystem.

\subsubsection{Entanglement breaking}
\label{ent_break}

A useful property to determine environmentally induced decoherence
is whether it destroys any entanglement the primary subsystem $Q$
has with an external subsystem. The superoperator $\Lambda$ (see
appendix \ref{QOFandJI}) describing the effect of the interaction is
`entanglement breaking' \cite{PhysRevA.71.032350} if, when acting on
a subsystem $b$ of dimension $d_b$, the final state
$\rho_{ab}^{fin}=\openone\otimes\Lambda_a(\rho_{ab}^{ini})$ is
separable for every (possible entangled) initial state
$\rho_{ab}^{ini}$ of the composite system of $b$ and an external
subsystem $a$ of dimension $d_a$ (not necessarily a copy of $b$).
Importantly, when the state $\rho^\Lambda$ obtained as described in
appendix \ref{QOFandJI} is separable, the superoperator $\Lambda$
will be entanglement breaking since this implies the corresponding
quantum operation $\varepsilon$ has a Kraus representation composed
entirely of projectors. Separability can be determined, using the
positive partial transposition (PPT) criterion
\cite{PhysRevLett.77.1413, Horodecki19961}, by computing the minimum
eigenvalue $\mu (t)$ of the matrix $(\rho^\Lambda)^{T_a}$. The
induced decoherence will be entanglement breaking iff $\mu
(t)\geq0$.

\section{Results: Ising interaction}
\label{Ising}

In this section we present results for the single spin decoherence
induced by the Ising interaction with the LMG bath spins. Before
doing so, it is useful to discuss exactly how decoherence arises in
this system.

Consider the qubit initially prepared in a generic superposition of
its ground and excited states, such that the global state at time
$t=0$ is
\begin{equation}
|\Psi(0)\rangle = \big(a_0|0\rangle + a_1|1\rangle \big)\otimes
|GS\rangle \textrm{ ,}
\end{equation}
where the coefficients $a_0$ and $a_1$ satisfy $|a_0|^2+|a_1|^2=1$.
The subsequent evolution under $H_T$ for times $t>0$ is described by
\begin{eqnarray}
|\Psi(t)\rangle & = & a_0e^{-i\omega t}|0\rangle\otimes e^{-iH_+t}|GS\rangle \nonumber \\
& & +\ a_1e^{i\omega t}|1\rangle\otimes e^{-iH_-t}|GS\rangle
\textrm{ ,} \label{Psi(t)}
\end{eqnarray}
where $S_z^k=\sum_{i=1}^k{\sigma_i^z/2}$, and $H_\pm=H_R\pm
2\epsilon S_z^k/\sqrt{k}$ are perturbed bath Hamiltonians. In
general, the bath evolves to different states dependent on whether
the qubit was initially in the state $|0\rangle$ or $|1\rangle$, and
therefore Eq. (\ref{Psi(t)}) is an entangled state of the two
subsystems. After tracing out the bath we obtain the reduced density
matrix for the qubit (in the eigenbasis $\{|0\rangle,|1\rangle\}$)
\begin{equation}
\rho(t) = \textrm{Tr}_R\{ |\Psi(t)\rangle\langle\Psi(t)|\} =
\begin{pmatrix}
|a_0|^2 & \rho_{12}(t) \\
\rho^*_{12}(t) & |a_1|^2
\end{pmatrix} \textrm{,}
\label{rho_matrix}
\end{equation}
where
\begin{equation}
\rho_{12}(t) = a_0a_1^*e^{-2i\omega t}D^*(t) \textrm{ .}
\end{equation}
The relative populations of the ground and excited states of the
qubit are conserved, but dephasing noise is introduced through the
decoherence factor
\begin{equation}
D(t) = \langle GS|e^{iH_+t}e^{-iH_-t}|GS\rangle \textrm{ .}
\label{decoherence}
\end{equation}
Note that the induced decoherence is independent of $\omega$. The
purity of the state in Eq. (\ref{rho_matrix}) is
\begin{equation}
P[\rho(t)] = 1 - 2|a_0|^2|a_1|^2\left[1-L(t)\right] \textrm{ ,}
\end{equation}
where $L(t)\equiv |D(t)|^2$ is a real quantity taking the form of a
Loschmidt Echo (LE) \cite{Peres} for the bath. Unlike the purity,
$L(t)$ is independent of the initial state of the qubit, aside from
the fact that it is undefined for $a_0=0,1$. Thus, the average
purity is trivially related to the LE via $\overline{P(t)} =
[2+L(t)]/3$.

By the term Loschmidt Echo we mean that $L(t)$ quantifies the
distance between the two bath states that have evolved from the GS
under the two perturbed bath Hamiltonians $H_\pm$. When the distance
between these states is small, the entanglement between the qubit
and the bath is low and the average purity is close to one.
Conversely, when the LE is zero the two time-evolved bath states are
orthogonal and the qubit is maximally entangled with the bath. In
such cases, the qubit undergoes total dephasing to a completely
mixed state and the average purity takes its minimum value of 2/3.
Also, this type of dephasing noise is entanglement breaking iff
$L(t)=0$ \cite{PhysRevA.71.032350}.

A useful quantity to consider that will give insight into how the
decoherence varies with link number $k$ is the Fourier Transform of
the decoherence factor in Eq. (\ref{decoherence}), given by
\begin{equation}
\chi(E) = \sum_{i,j}{\langle
GS|\phi^+_i\rangle\langle\phi^+_i|\phi^-_j\rangle\langle\phi^-_j|GS\rangle\delta(E-E^-_j+E^+_i)}
\textrm{ ,} \label{chi}
\end{equation}
where $|\phi^\pm_i\rangle$ are eigenvectors of the perturbed bath
Hamiltonians $H_\pm$, and $E^\pm_i$ their eigenenergies.
Contributions to $\chi(E)$ occur at energy differences $E^-_j-E^+_i$
corresponding to eigenstates $|\phi^-_j\rangle$ and
$|\phi^+_i\rangle$ that simulaneously connect to (i.e. are not
distant from) each other and the bath GS.

\subsection{Single link, $k=1$}

The simplest case to consider first is that of a single link between
the qubit and the bath. In particular, for an isotropic bath
($\gamma=1$) the GS in the normal phase ($\lambda<h$) is the
spin-polarized state $|N/2,N,2\rangle$, which is an eigenstate of
both perturbed bath Hamiltonians $H_\pm$. The bath remains in the GS
for the entirety of the evolution and simply induces an additional
$z$-rotation on the initial qubit state through an angle $2\epsilon
t$, thus there is zero dephasing.

In the broken phase ($\lambda>h$) we observe a loss of qubit
coherence and find that the average purity has an oscillatory
behavior with time, as shown in Fig. \ref{Ising_single}(a). For an
isotropic bath, we show in appendix \ref{appendix_k=1} that the GS
in the broken phase decomposes into two terms when we move to the
bipartite subspace
$\mathbb{S}^1_{1/2}\otimes\mathbb{S}^{N-1}_{(N-1)/2}$. The evolution
of the bath under $H_\pm$ is further confined to a subspace
comprising only the two states of the decomposition, allowing us to
describe the bath propagators $e^{-iH_\pm t}$ by 2 x 2 matrices (see
Eqs. (\ref{eH_- action}) and (\ref{eH_+ action})). The resulting
expression for the purity is too complicated to be given explicitly,
however, we can extract the salient behavior from the form of the
two propagators. The off-diagonal matrix elements of $e^{-iH_- t}$
vary sinusoidally such that it periodically becomes equal to the
identity matrix with a frequency
$\eta_-=2\sqrt{\lambda^2+\epsilon(\epsilon+2h)}$ that is independent
of bath size $N$. Similarly, $e^{-iH_- t}$ varies periodically with
a frequency $\eta_+=2\sqrt{\lambda^2+\epsilon(\epsilon-2h)}$. When
both propagators are the identity at the same time, the LE becomes
unity and there is a revival of full qubit coherence. This
`coherence time' $\tau_c$ must simultaneously satisfy
$\eta_-\tau_c=2l_-\pi$ and $\eta_+\tau_c=2l_+\pi$, where $l_-$ and
$l_+$ are integers. In general, we observe fast oscillations of the
average purity determined by the higher frequency $\eta_-$, which
are contained within an envelope function that is periodic with a
frequency $\eta_+$. High average purities of
$\overline{P(t,\lambda)}>0.98$ are obtained at times $t=l\tau_r$
($l$ an integer), where $\tau_r=\pi/\eta_+$ is the approximate
`rephasing time' corresponding to the first maximum of the envelope
function. This is plotted in Fig. \ref{Ising_single}(b) for various
values of the interaction strength $\epsilon$. In the limit
$\epsilon\ll h$, the frequencies $\eta_\pm$ are approximately equal
and a full revival of qubit coherence occurs at a time
$\tau_c\sim\tau_r=\pi/\lambda$. For $\epsilon=h$ the rephasing time
is asymptotic at $\lambda=h$, and thus its sensitivity to small
changes in the magnetic field allows for an observation of
criticality. However, criticality should be avoided if one wishes to
use the system for storage of the qubit state so that rephasing can
occur on short-time scales.

The behavior described above for the coupling to an isotropic bath
is independent of bath size $N$. Thus, the revivals of qubit
coherence are not a finite size effect and occur in the
thermodynamical limit. This independence with $N$ arises because the
bath Hamiltonian takes a particularly simple form for $\gamma=1$
that is diagonal in the eigenbasis spanned by Dicke states
$|S,M\rangle$, ultimately restricting the evolution to just a
two-dimensional subspace. As a result, the decoherence factor has a
maximum of four Fourier components, as shown in Fig.
\ref{Ising_single}(c); there is only a single component in the
normal phase where no dephasing occurs. Interestingly, when we
consider anisotropic baths such that the evolution utilizes the
entire $\mathbb{S}^1_{1/2}\otimes\mathbb{S}^{N-1}_{(N-1)/2}$
subspace, we find there are still only $n=4$ significant Fourier
components to the decoherence factor. Significant Fourier components are those of greatest amplitude that reproduce $98\%$ of the signal. Also, we observe a rapid convergence of the average
purity to the isotropic bath result with increasing bath size. In
Fig. \ref{Ising_single}(d) we have plotted the average purity across
criticality at time $t=5/h$ for various bath anisotropies and a bath
of $N=100$ spins. Away from criticality, the average purity for all
bath anisotropies is within $5\%$ of the isotropic result for a bath
of this size. Close to criticality the convergence is slower, and we
observe oscillations in the average purity in the normal phase that
disappear in the thermodynamical limit.

\begin{figure}
\includegraphics[width=4.1cm]{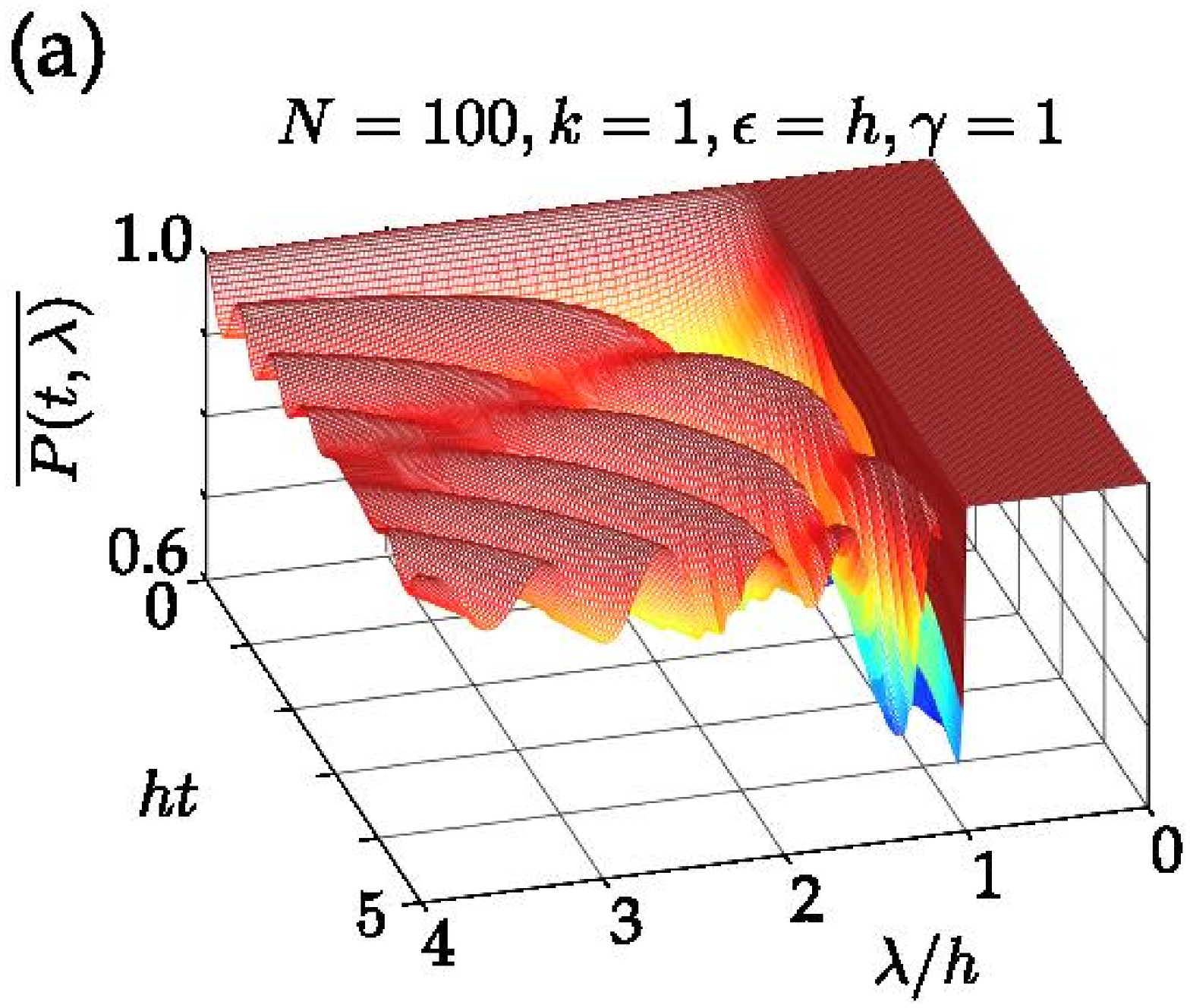}\hspace{0.25cm}
\includegraphics[width=4.1cm]{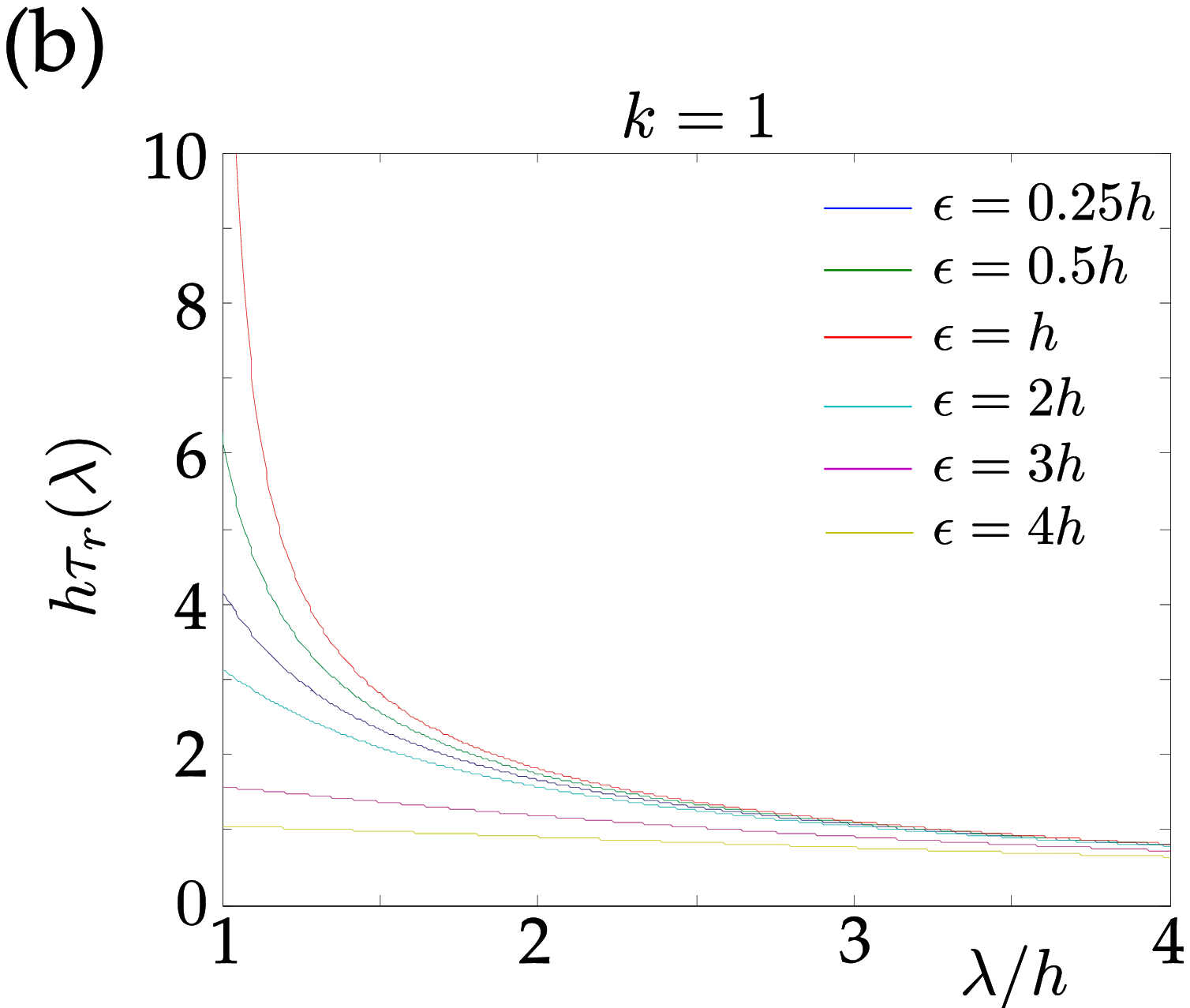}\vspace{0.3cm}
\includegraphics[width=4.1cm]{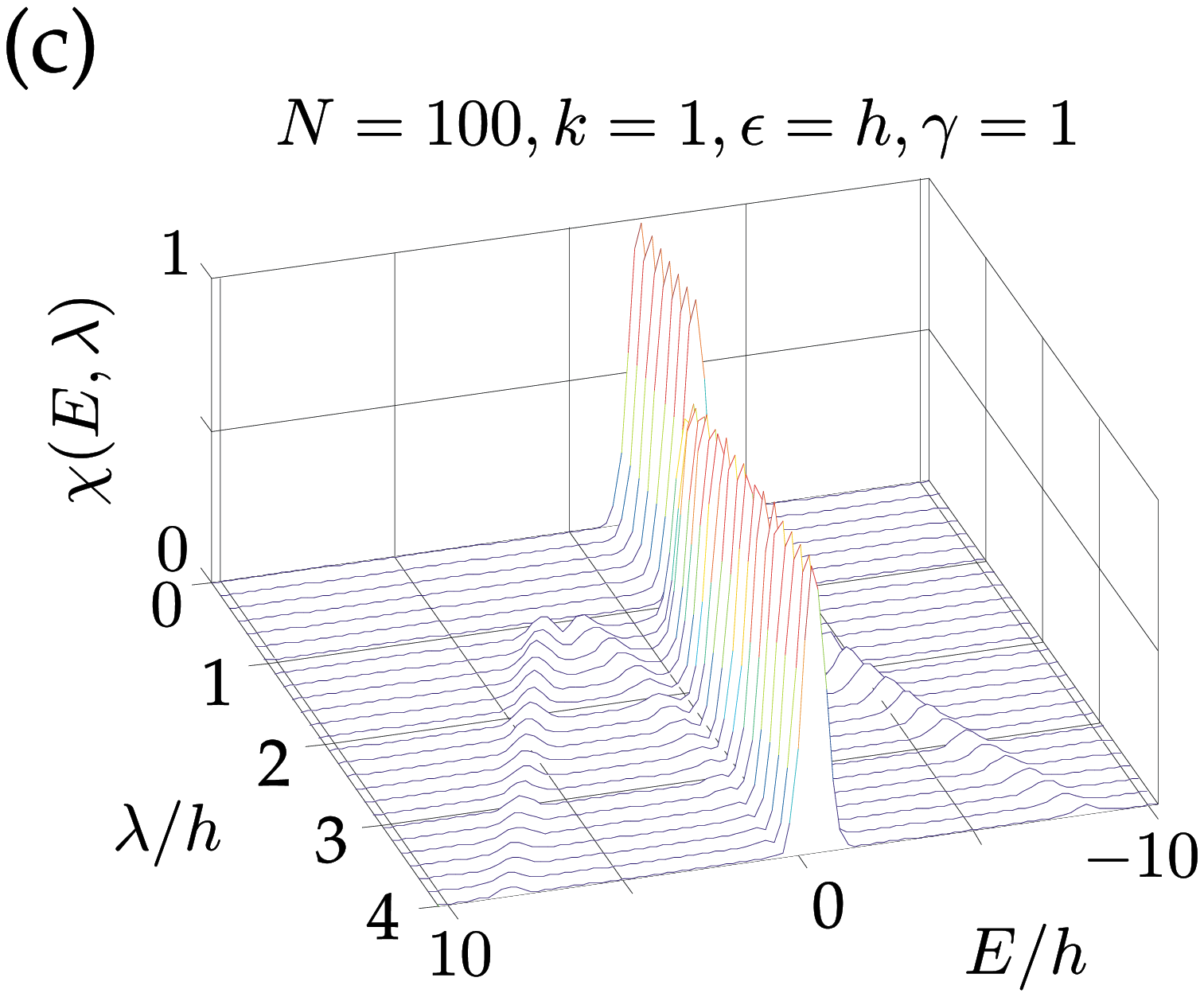}\hspace{0.25cm}
\includegraphics[width=4.1cm]{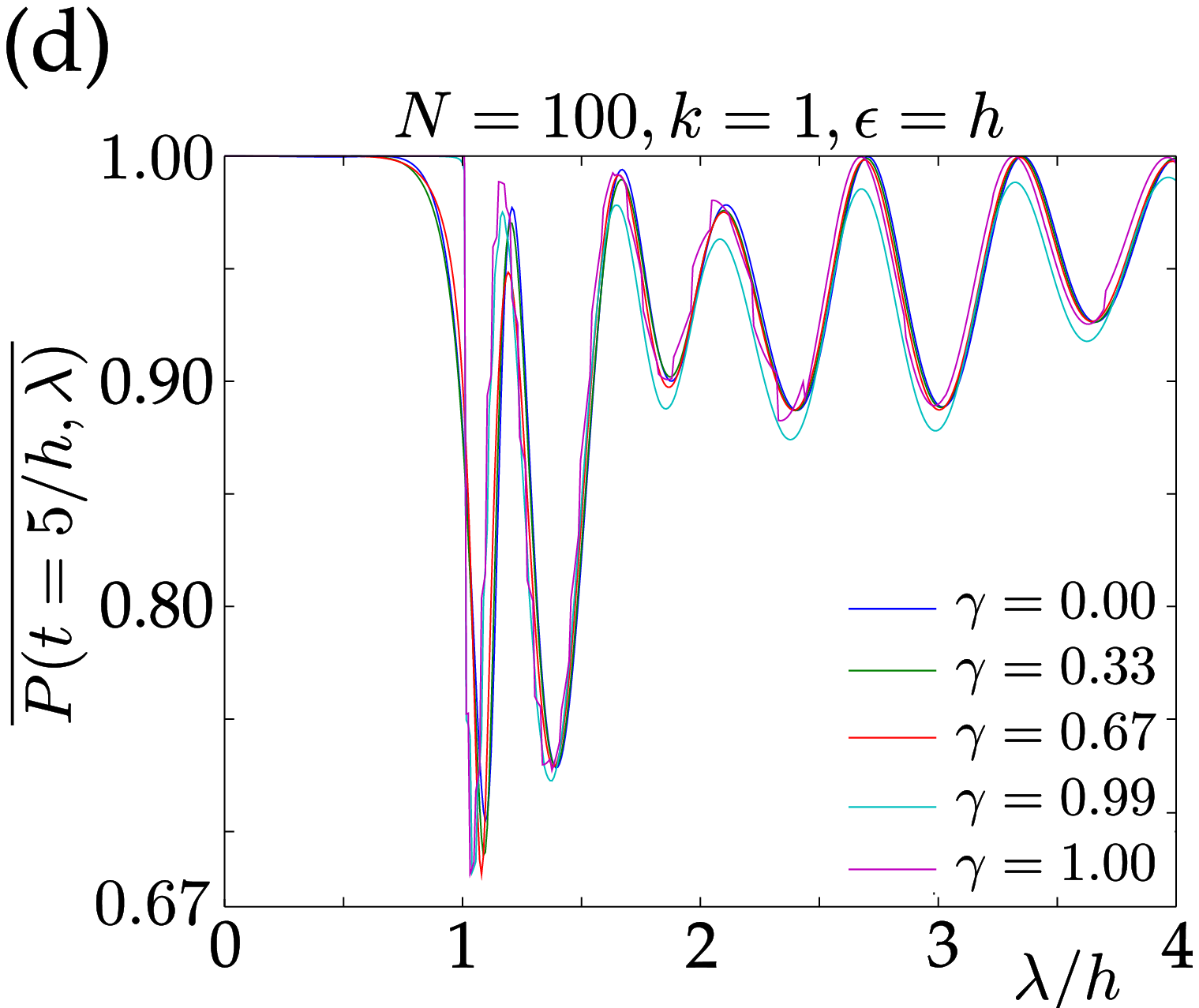}
\caption{Ising interaction, single link ($k=1$): (a) the average
purity plotted against $\lambda$ and $t$. (b) The approximate
rephasing time $\tau_r$ plotted against $\lambda$ for various values
of $\epsilon$. (c) The Fourier Transform of the decoherence factor
as a function of energy and $\lambda$. We have used Gaussian
envelopes at each Fourier component for ease of viewing. (d) The
average purity plotted against $\lambda$, at $t=5/h$, for various
values of $\gamma$.} \label{Ising_single}
\end{figure}

\subsection{Completely connected, $k=N$}
\label{Ising_k=N}

We now consider the opposite extreme where the qubit is coupled to
all spins in the bath, i.e. the central spin model. For an
isotropic bath the average purity is trivially equal to one for all
$\lambda$; the interaction term commutes with the bath Hamiltonian
and thus the qubit and the bath cannot become entangled. The qubit
experiences a $z$-rotation through an angle $\epsilon Nt$ in
addition to that of its free evolution \footnote{The scaling with
bath size $N$ of the induced phase may make it difficult for one to
keep track of this for large baths.}.

\begin{figure}
\includegraphics[width=4.1cm]{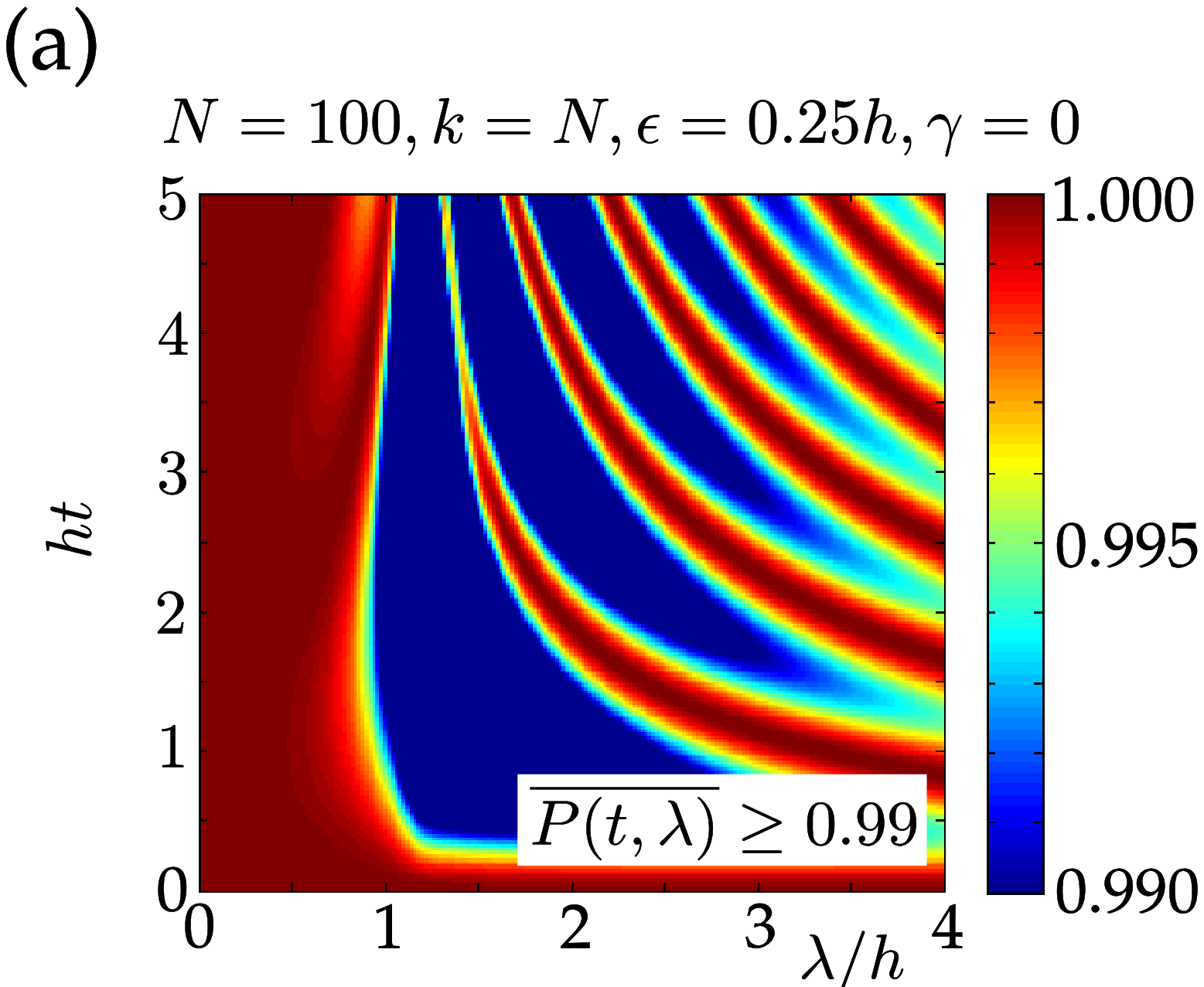}\hspace{0.25cm}
\includegraphics[width=4.1cm]{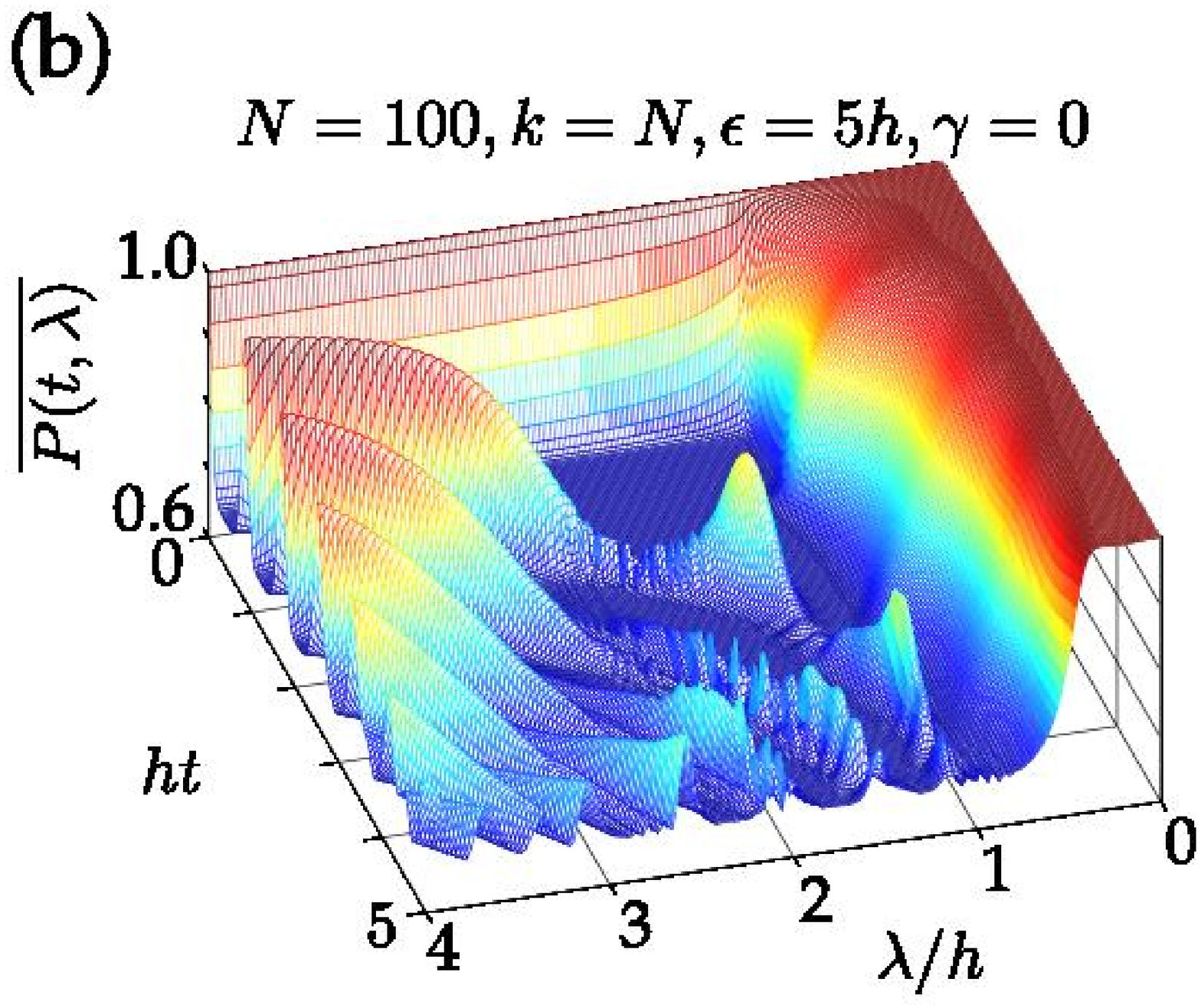}\vspace{0.3cm}
\includegraphics[width=4.1cm]{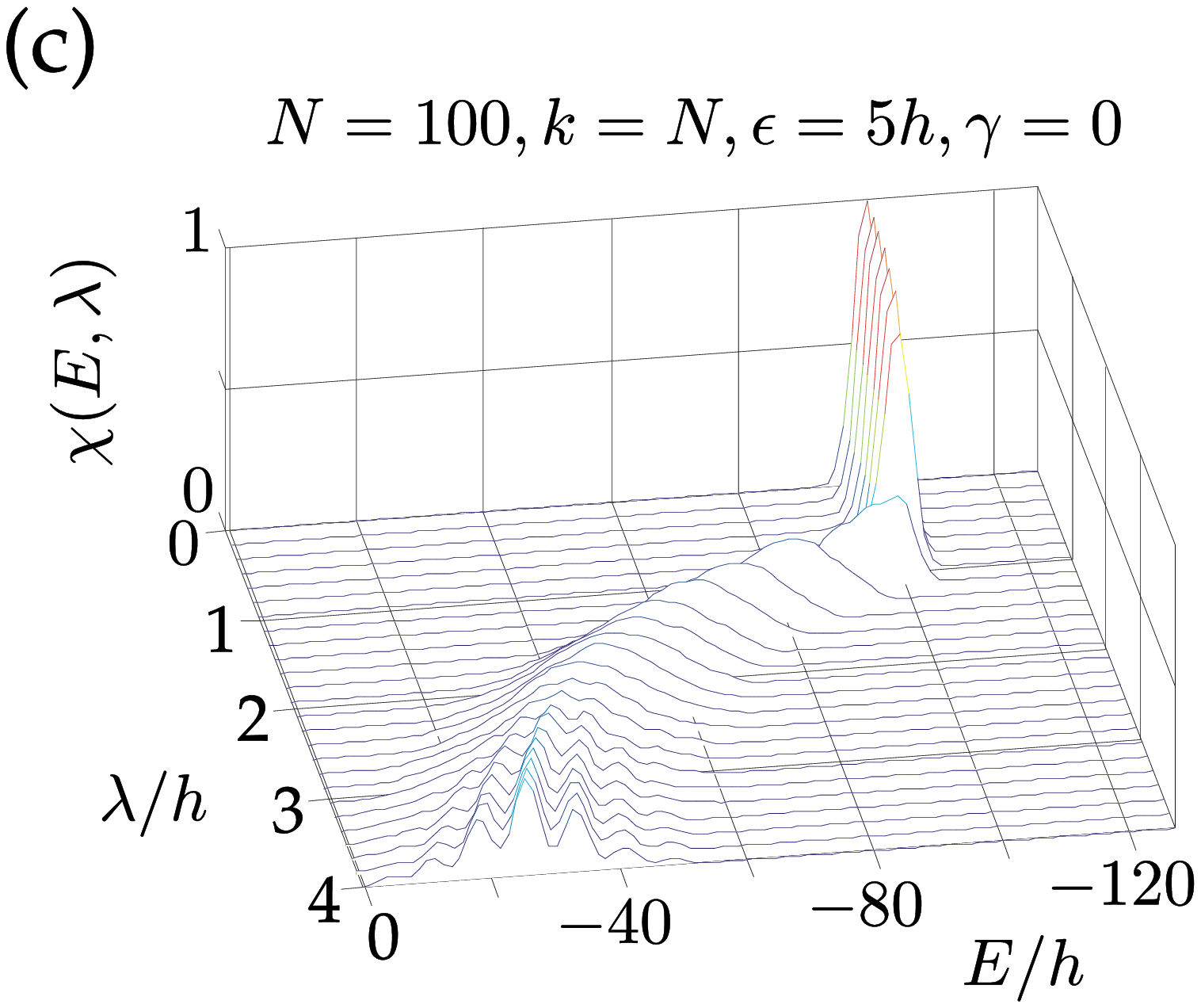}\hspace{0.25cm}
\includegraphics[width=4.1cm]{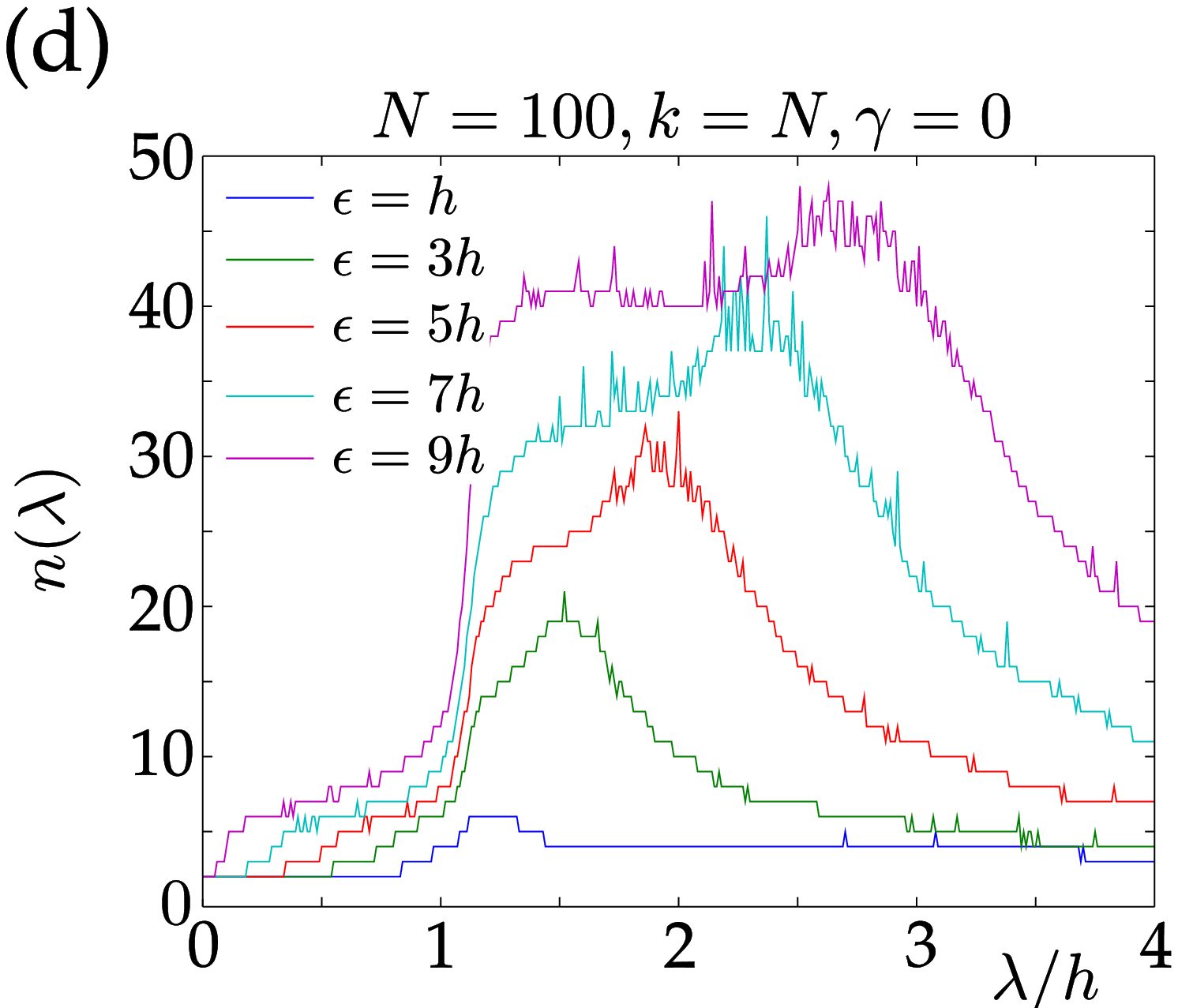}
\caption{Ising interaction, completely connected ($k=N$): (a) a
colour map of the average purity plotted against $\lambda$ and $t$
for $\epsilon=0.25h$. The map has a threshold at $\overline{P}=0.99$
to improve clarity. (b) The average purity plotted against $\lambda$
and $t$ for $\epsilon=5h$. (c) The Fourier transform of the
decoherence factor as a function of energy and $\lambda$ for
$\epsilon=5h$; again, we have used Gaussian envelopes at each
Fourier component. (d) The number of Fourier components $n$ of
greatest amplitude that reproduce $98\%$ of the signal, plotted
against $\lambda$, and for various values of $\epsilon$.}
\label{Ising_N}
\end{figure}

Figure \ref{Ising_N}(a) shows a color map of the average purity
against $\lambda$ and $t$ for an anisotropic bath ($\gamma=0$) with
$\epsilon=0.25h$. At this low interaction strength the average
purity is qualitatively similar to the single link scenario; there
is almost zero dephasing in the normal phase, whilst oscillations
occur within an envelope function in the broken phase. The rephasing
time $\tau_r$ is asymptotic at criticality but contrary to the
single link scenario we find it is independent of $\epsilon$. As
would be expected for this description the Fourier Transform of the
decoherence factor for low interaction strengths is qualitatively
similar to that shown in Fig. \ref{Ising_single}(c); there is one
Fourier component with a large amplitude accompanied by a few
($n<15$) of smaller amplitude.

As the interaction strength is increased above $\epsilon=h$ the
behavior becomes markedly different. The maxima in the average
purity are suppressed and, as shown in Fig. \ref{Ising_N}(b) for
$\epsilon=5h$, high average purities are no longer achieved on
short-time scales, except in the limit $\lambda\rightarrow 0$. In
addition, away from the maxima the LE is close to zero indicating
that the evolution is almost entanglement breaking. The Fourier
Transform of the decoherence factor for $\epsilon=5h$, shown in Fig.
\ref{Ising_N}(c), contains a large number $n$ of significant
components but no single one of high amplitude, resulting in the
suppression of the maxima. In Fig. \ref{Ising_N}(d) we have plotted
$n$ across criticality for various values of $\epsilon$. The number
of significant Fourier components increases rapidly as we move from
the normal to the broken phase, and contrary to the single link
scenario increases at all intra-bath coupling strengths with
$\epsilon$.

Interestingly, for any $\epsilon$ the energy spread and number of
components rapidly converge to finite values with increasing bath
size. Thus, although at high interaction strengths there is a decay
of qubit coherence on short-time scales, full revivals will
eventually occur for any $\epsilon$ even in the thermodynamical
limit, with the coherence time $\tau_c$ independent of $N$ for large
baths, $N\gtrsim 100$. Finally, we note that the behavior described
in this subsection is only qualitatively the same for different bath
anisotropies and does not tend to the isotropic behavior in the
limit $\gamma\rightarrow 1$.

\subsection{Multiple links}
\label{Ising_multiple}

In the previous two subsections we have seen approximate rephasing
occurs on short-time scales for any interaction strength when the
qubit is connected via a single link to the bath, whilst for a
completely connected qubit rephasing is suppressed at high
interaction strengths. We now examine how this transition occurs as
we gradually increase the number of links from the scenario $k=1$ to
$k=N$. We consider an intra-bath coupling strength of $\lambda=3h$
such that we are away from criticality and rephasing is possible on
short-time scales for any link number.

\begin{figure}
\includegraphics[width=4.1cm]{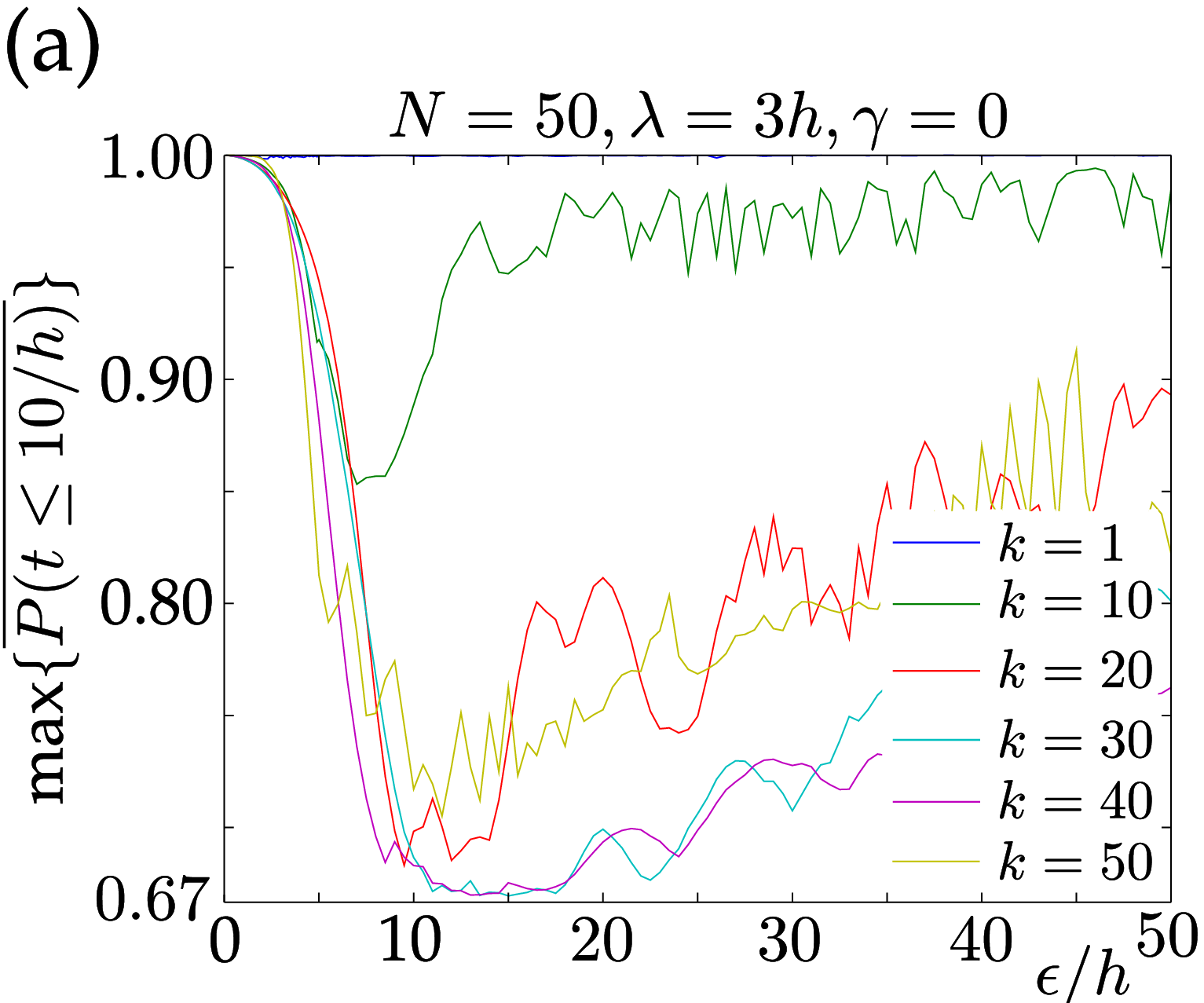}\hspace{0.25cm}
\includegraphics[width=4.1cm]{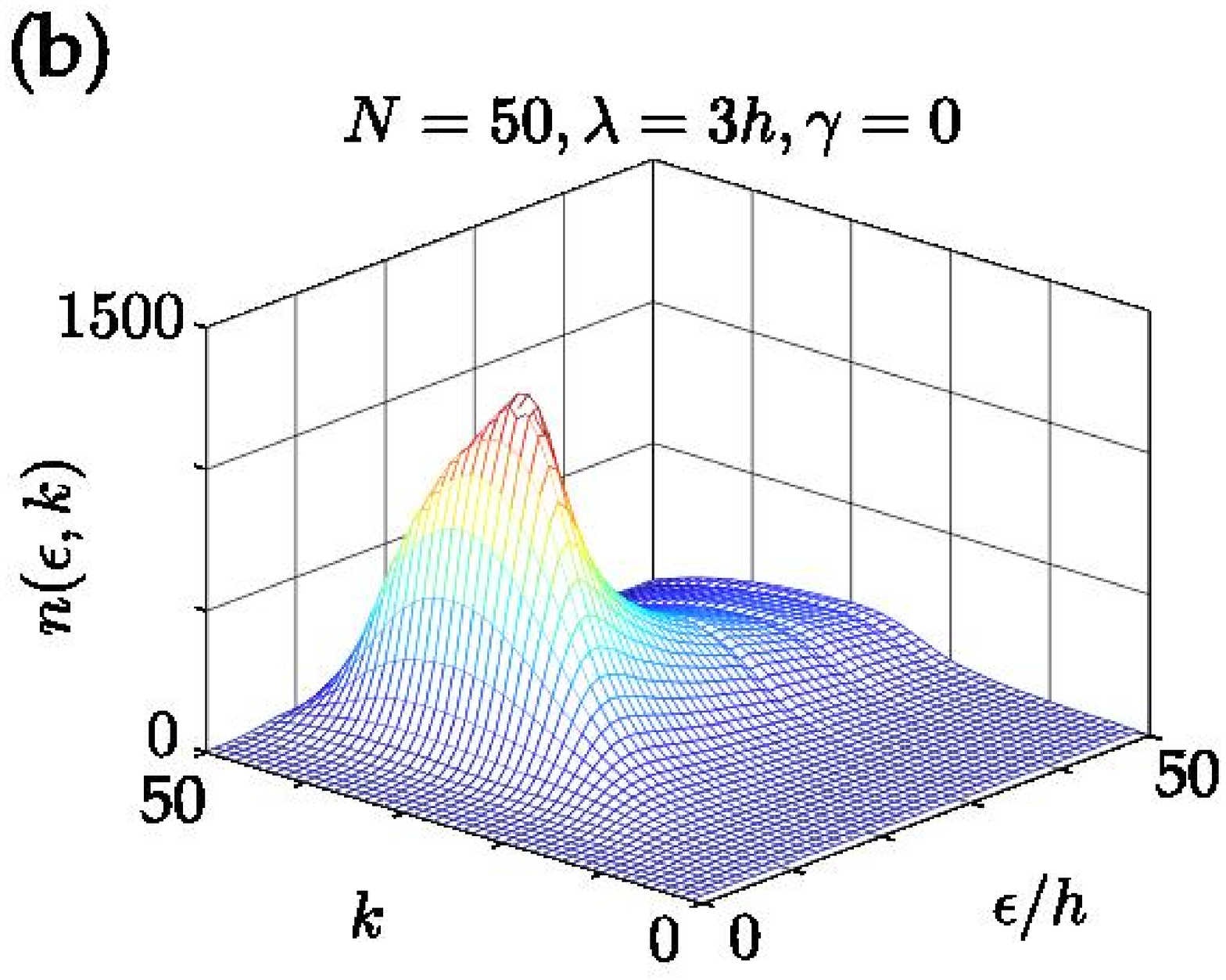}\vspace{0.3cm}
\includegraphics[width=4.1cm]{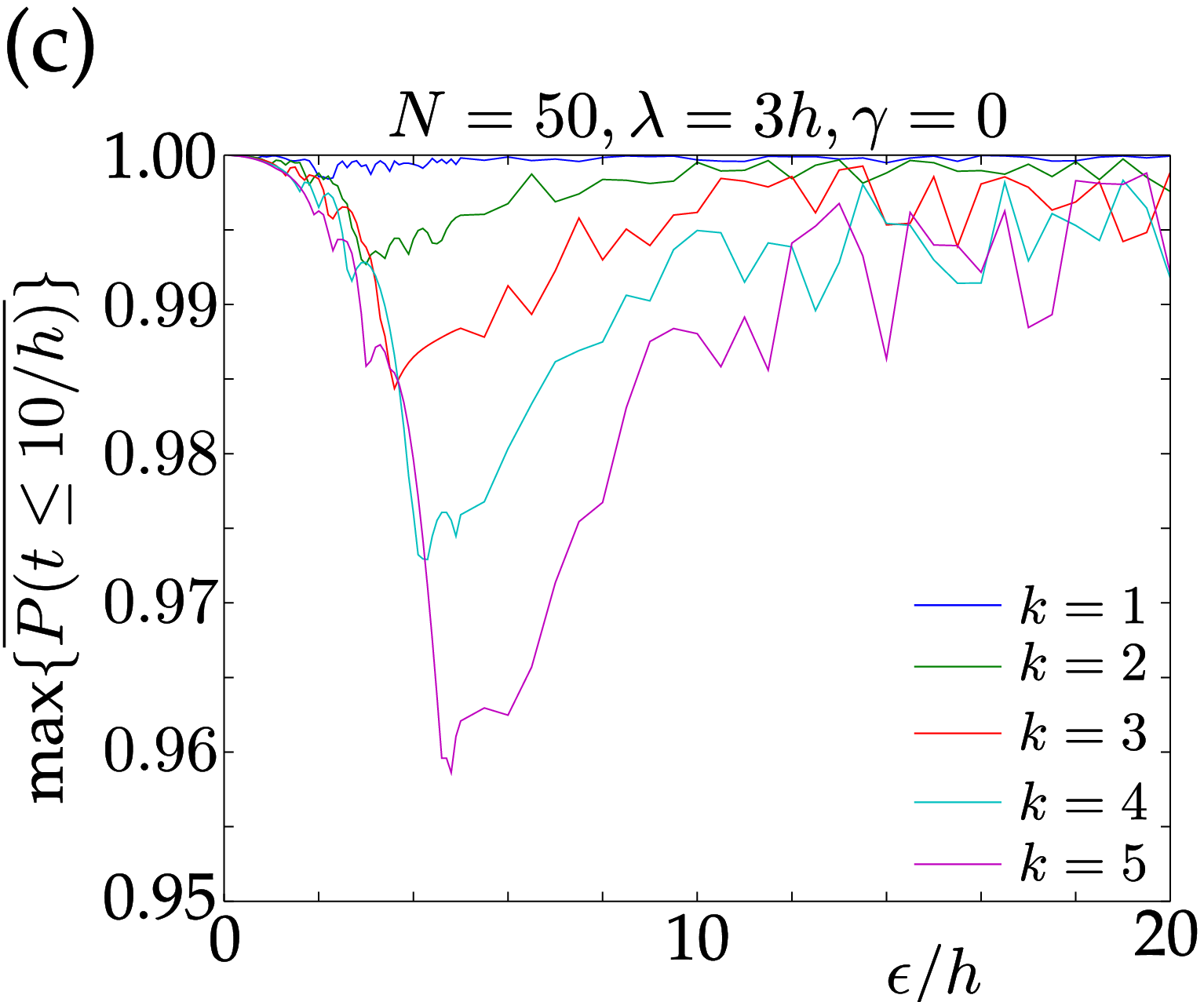}\hspace{0.25cm}
\includegraphics[width=4.1cm]{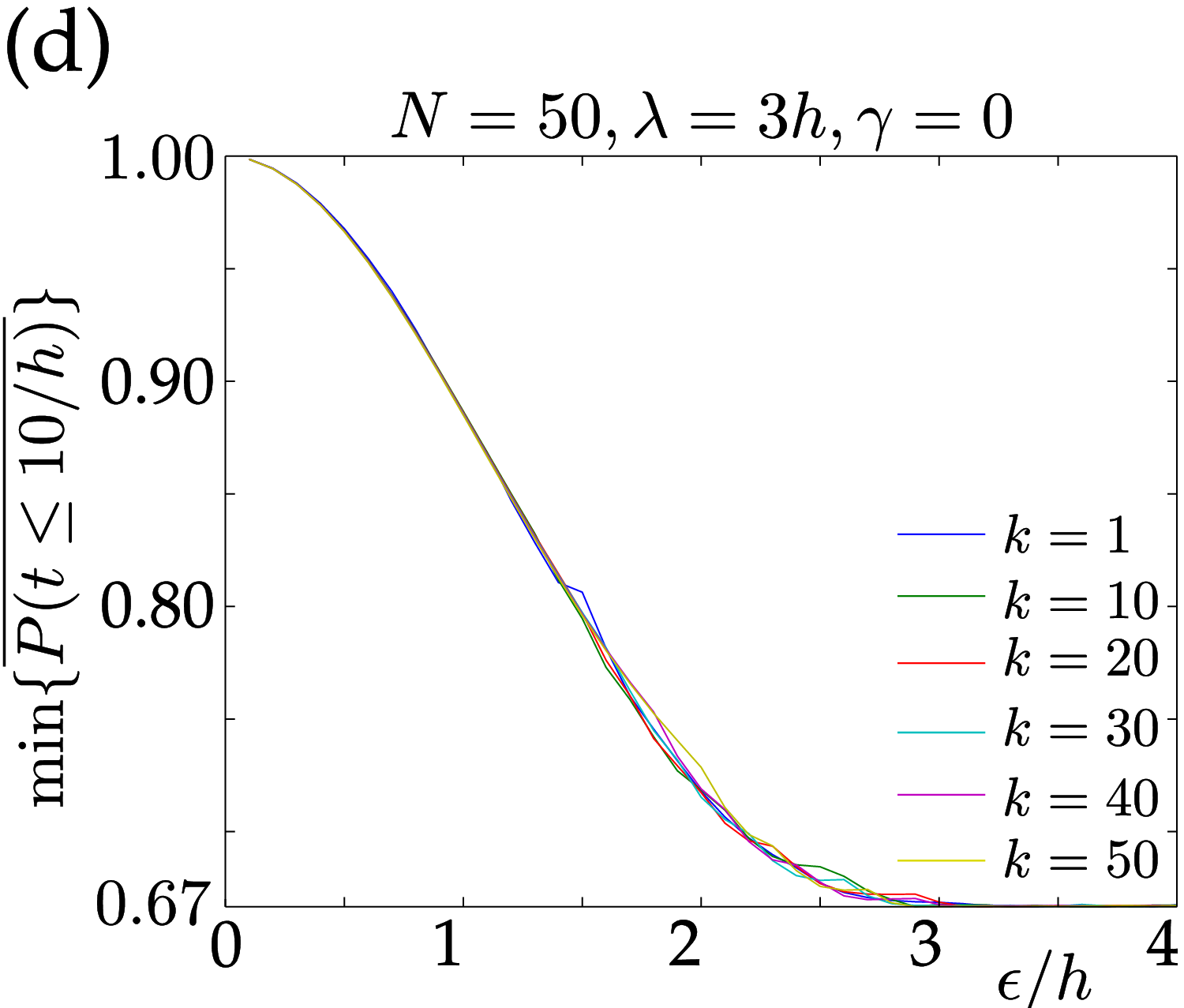}
\caption{Ising interaction: (a) the maximum value of the Loschmidt
Echo at $\lambda=3$ for rephasing within a time $t\leq 10/h$, as a
function of $\epsilon$, and for a range of $k$. (b) The number of
Fourier components $n$ of greatest amplitude that reproduce $98\%$
of the signal as a function of $\epsilon$ and $k$. (c) same as (a)
but for only a few links. (d) The minimum value of the Loschmidt
Echo within $t\leq 10/h$ at $\lambda=3$, plotted against
$\epsilon$.} \label{Ising_k}
\end{figure}

In Figs. \ref{Ising_k}(a) and \ref{Ising_k}(c) we have plotted the
maximum value of the average purity for rephasing within a time
$t\leq 10/h$, as a function of the interaction strength $\epsilon$,
and for a range of link numbers $k$. The bath size is $N=50$. High
average purities of $\overline{P}>0.98$ are achieved for any link
number below $\epsilon\sim 4h$, whilst above this interaction
strength a high average purity is only guaranteed for $k<3$. At
higher link numbers we observe the suppression of maxima up to
$\epsilon\sim 10h$ as discussed in Sec. \ref{Ising_k=N} for a
completely connected bath. Interestingly, above $\epsilon=10h$ the
maxima in the average purity increase back towards one, with the
rate of increase slowest for link numbers $k\sim30$-$40$. This
behavior is prevalent when we consider the number of significant
Fourier components $n$ as a function of link number and interaction
strength, shown in Fig. \ref{Ising_N}(b). For a few link numbers, we
have $n<10$ for all $\epsilon$ allowing approximate rephasing to
occur. For higher link numbers, the number of components increase
with $\epsilon$ up to a maximum value at $\epsilon\sim 10h$, before
decreasing once more at higher interaction strengths. The number of
components also has a maximum at $k\sim 35$ and therefore decreases
with large $k$, but albeit is still relatively large for a
completely connected qubit.

The restriction of the number of significant Fourier components can
be explained by considering the form of the perturbed bath
Hamiltonians $H_\pm=H_R\pm 2\epsilon S_z^k/\sqrt{k}$. At low
interaction strengths $\epsilon<h$, the perturbing terms are small
and the eigenstates of the two Hamiltonians are similar. There are
approximately $(k+1)(N-k+1)$ terms $\langle\phi_i^+|\phi_j^-\rangle$
that are non-zero, but only a few of these will connect to the GS
and thus $\chi(E)$ will have only a few components. As the
interaction strength is increased above $\epsilon=h$, the two sets
of eigenstates differ meaning that there are now $(k+1)^2(N-k+1)^2$
non-zero-terms $\langle\phi_i^+|\phi_j^-\rangle$, each of which will
be of smaller magnitude. There is a larger set of pairs of
eigenstates to connect to the GS leading to many more components.
When the interaction strength is increased further such that the
second terms dominate in $H_\pm$, the eigenstates again become
similar and the number of components asymptotically decreases to one
in the limit $\epsilon\rightarrow\infty$. The maximum in the number
of components with link number arises because the subspace
$\mathbb{S}^k_{k/2}\otimes\mathbb{S}^{N-k}_{(N-k)/2}$ is largest for
$k=N/2$, however the maximum is skewed towards $k=N$ so that a
completely connected qubit still has a relatively large number of
components compared to the case $k=1$.

Finally, we note that although the maxima in the average purity
depend on link number, the minima do not, as shown in figure
\ref{Ising_k}(d). The minima decrease towards, but do not reach, the
minimum value of 2/3 (corresponding to an LE of zero) with
increasing interaction strength and so the interaction is never
formally entanglement breaking.

\section{Results: LMG interaction}
\label{LMG}

We now present the results for the qubit interacting with the bath
spins via the dissipative LMG interaction. Contrary to the Ising
interaction, the relative populations of the ground and excited
states of the qubit can vary in time in addition to the coherence.
This means the average purity can take a minimum value of 1/2, but
more significantly it is now dependent on the energy difference
$\omega$ between the two levels of the qubit. In the following, we
consider the scenario \footnote{We note that in this scenario the
qubit cannot be spectroscopically isolated from the bath ensemble.}
in which the qubit is subject to an identical external field as the
bath spins, thereby setting $\omega=h$. In tests where $\omega$ was
varied decoherence was suppressed in all cases as $\omega$ was
increased, thus the scenario considered is in the worst-case regime.

One important consideration for the LMG interaction, in contrast to
the Ising interaction, is that the induced decoherence is
non-trivially dependent on the initial qubit state. This means that
the purity for a given initial state can have significantly
different behavior to the average purity, and indeed can be unity
when the average purity is not. It is interesting in the following
results that we observe the average purity oscillating away from and
back to unity, indiciating universal rephasing at certain times for
any given initial qubit state. The evolution is also entanglement
breaking for certain periods, which is shown in the following
results using the quantity $\mu(t)$ discussed in Sec.
\ref{ent_break}.

\subsection{Single link, $k=1$}

\begin{figure}
\includegraphics[width=4.1cm]{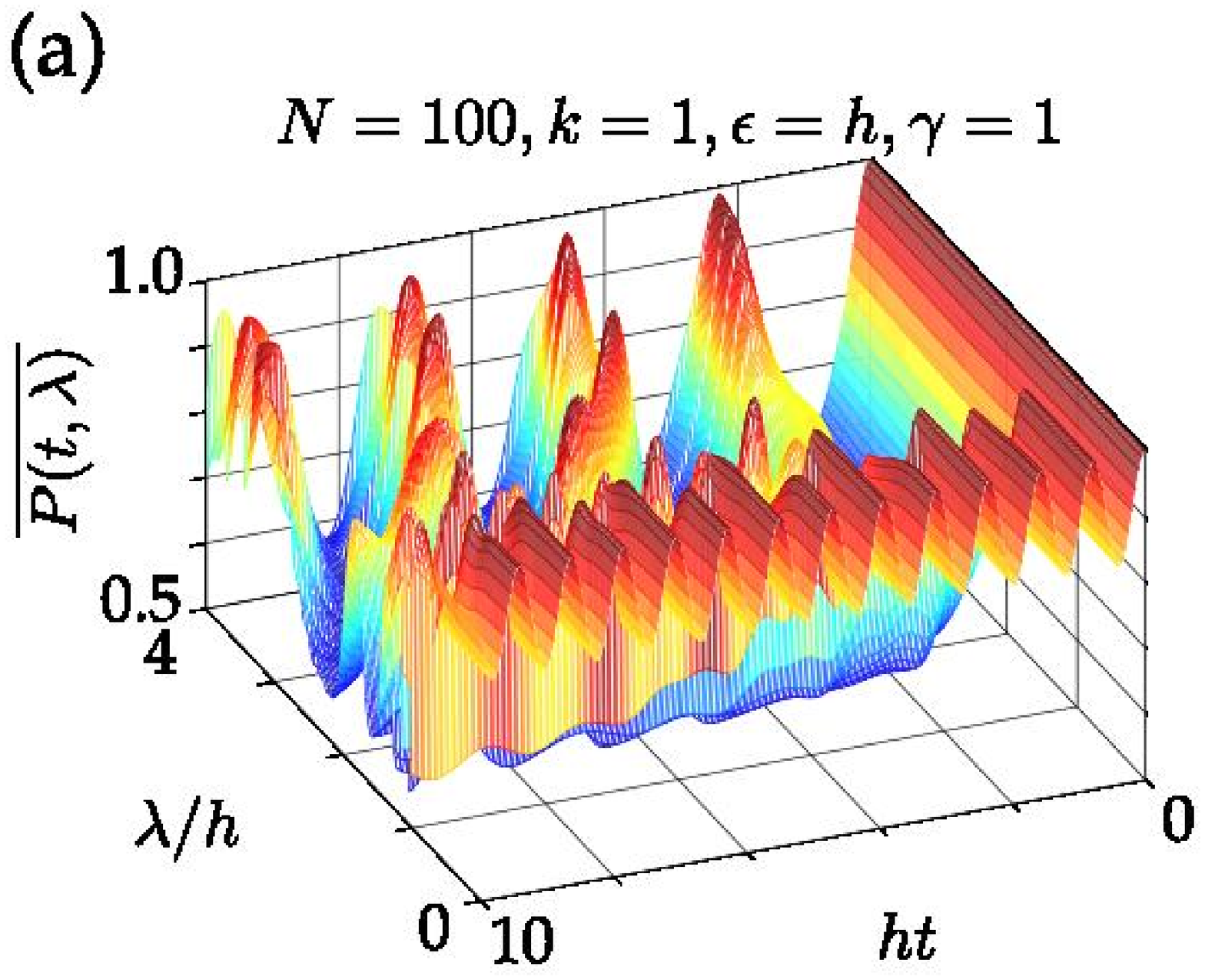}\hspace{0.25cm}
\includegraphics[width=4.1cm]{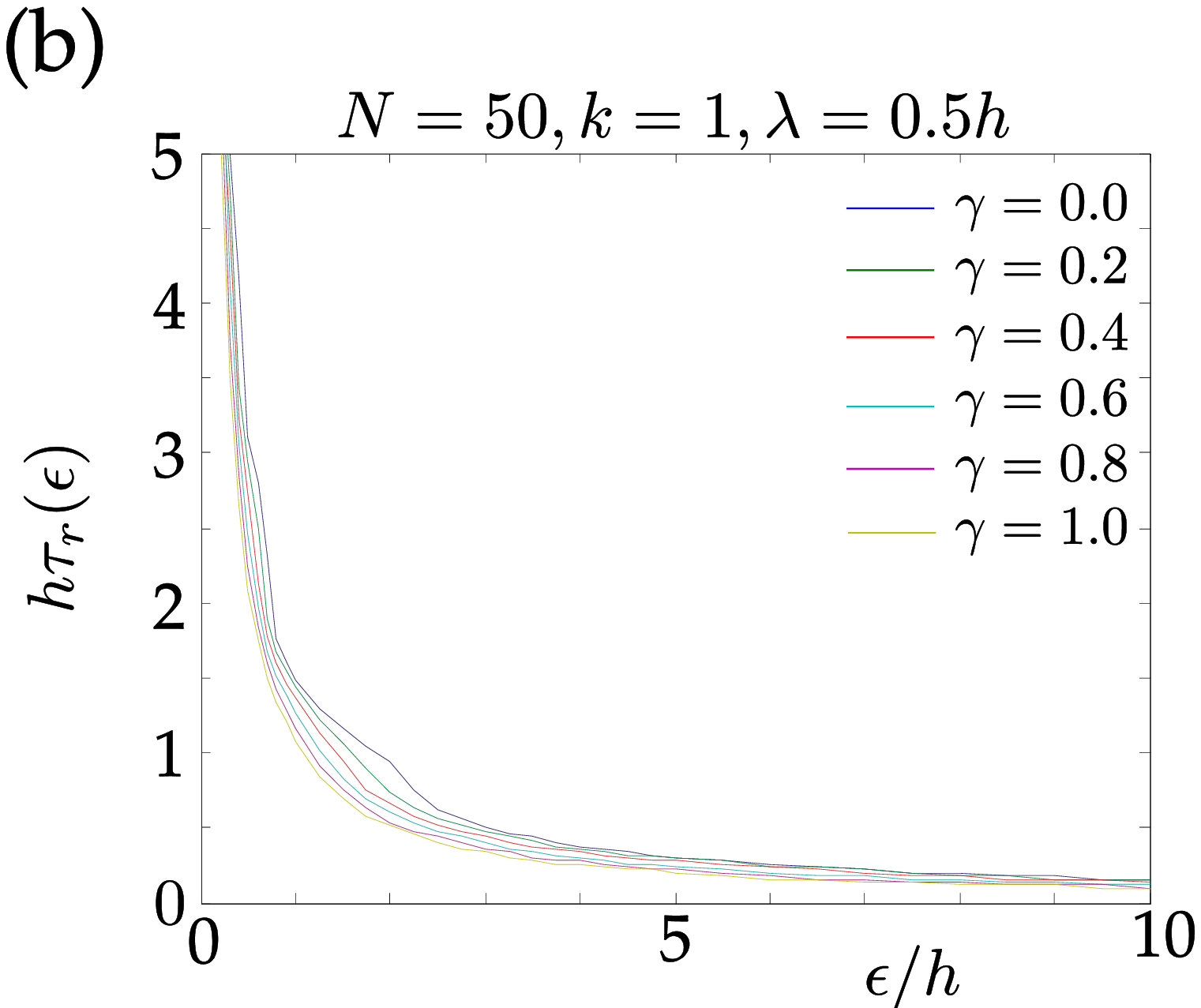}\vspace{0.3cm}
\includegraphics[width=4.1cm]{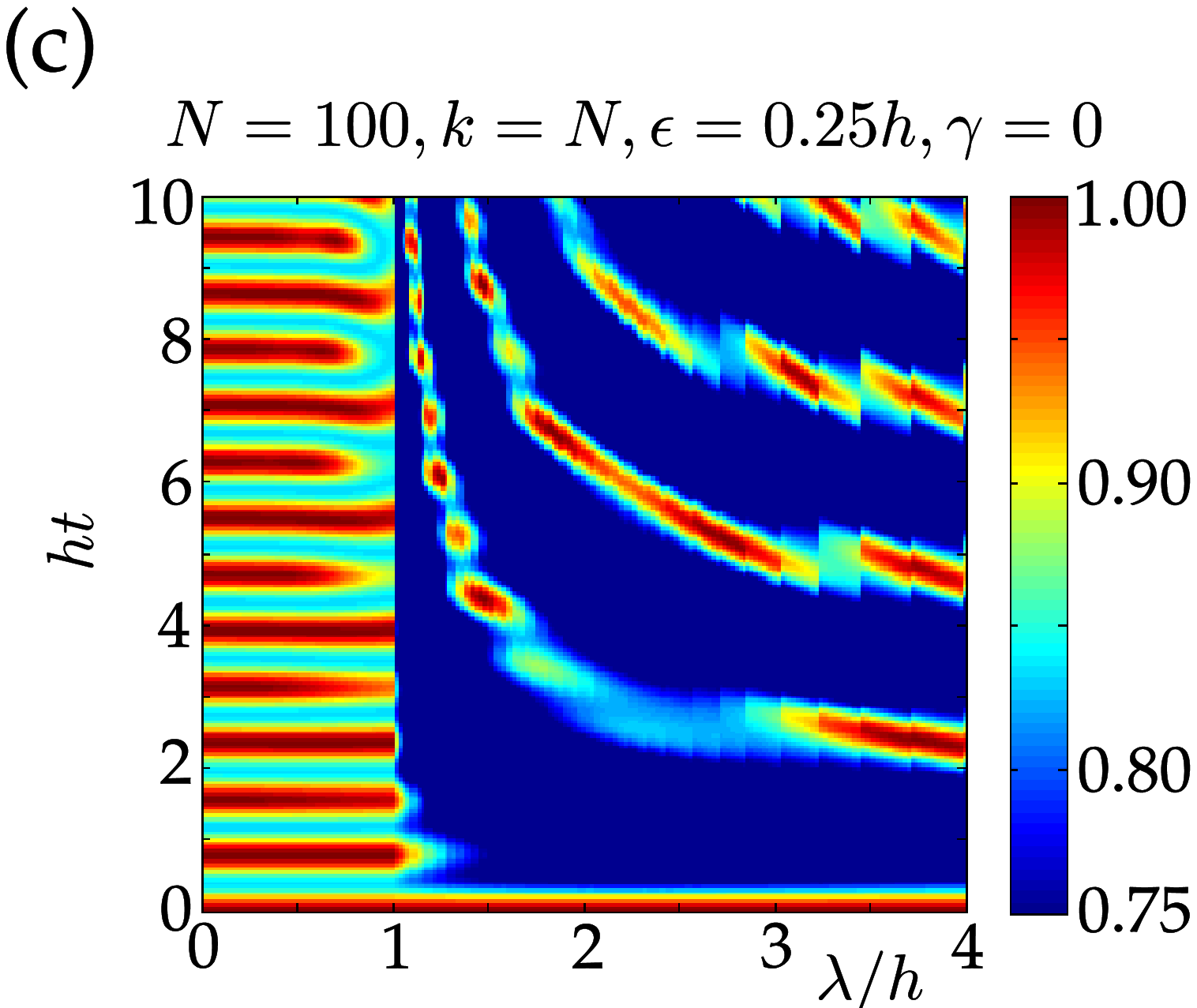}\hspace{0.25cm}
\includegraphics[width=4.1cm]{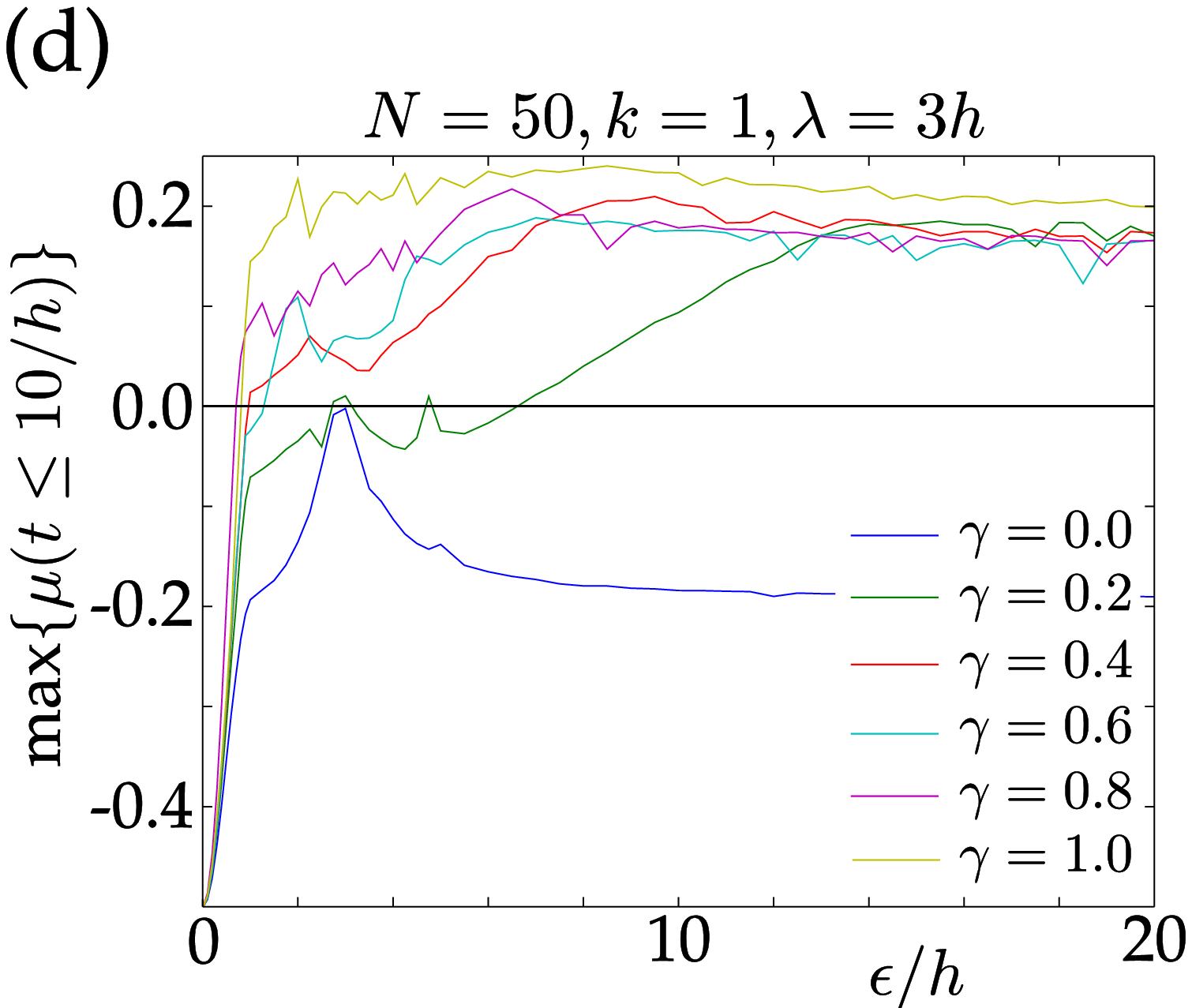}
\caption{LMG interaction, single link ($k=1$): (a) the average
purity plotted against $\lambda$ and $t$ for $\gamma=1$. (b) The
rephasing time $\tau_r$ for the bath in the normal phase
($\lambda=0.5$), as a function of $\epsilon$ and for various values
of $\gamma$. (c) A color map of the average purity plotted against
$\lambda$ and $t$ for $\epsilon=h$. The map has a threshold at
$\overline{P}=0.75$ to improve clarity. (d) The maximum value of
$\mu (t)$ within $t\leq 10/h$ at $\lambda=3h$, as a function of
$\epsilon$ and for various values of $\gamma$. Values above zero
indicate that the evolution is entanglement breaking.}
\label{lmg_single}
\end{figure}

In Fig. \ref{lmg_single}(a) we have plotted the average purity for
the qubit connected via a single link to an isotropic bath
($\gamma=1$) with an interaction strength $\epsilon=h$. For the LMG
interaction the average purity is not equal to unity in the normal
phase for an isotropic bath, but instead we observe approximate
revivals with a rephasing time $\tau_r$ that is independent of
intra-bath coupling strength $\lambda$. The rephasing time increases
rapidly with interaction strength and is almost independent of bath
anisotropy, as shown in Fig. \ref{lmg_single}(b). The interaction is
never entanglement breaking for any $\gamma$ or $\epsilon$ in this
phase.

As for the Ising interaction, we observe approximate revivals in the
broken phase for all $\gamma$ and $\epsilon$. The rephasing time
$\tau_r$ behaves in a similar fashion to the Ising case with a
single link: it is almost independent of $\gamma$; increases with
interaction strength up to $\epsilon=h$ where it is asymptotic at
criticality (see Fig. \ref{lmg_single}(c)); and decreases at higher
interaction strengths. In Fig. \ref{lmg_single}(d) we have plotted
the maximum value of $\mu (t)$ within a time $t\leq 10/h$ for
various bath anisotropies. We observe that the induced decoherence
is periodically entanglement breaking in time for $\gamma>0.1$ and
interaction strengths $\epsilon>8h$.

For the LMG interaction we again find that the average purity is
independent of bath size for large $N$. Thus, not only would we
observe revivals in the thermodynamical limit, but interestingly
they will occur after periods when the induced decoherence is
entanglement breaking.

\subsection{Completely connected, $k=N$}

\begin{figure}
\includegraphics[width=4.1cm]{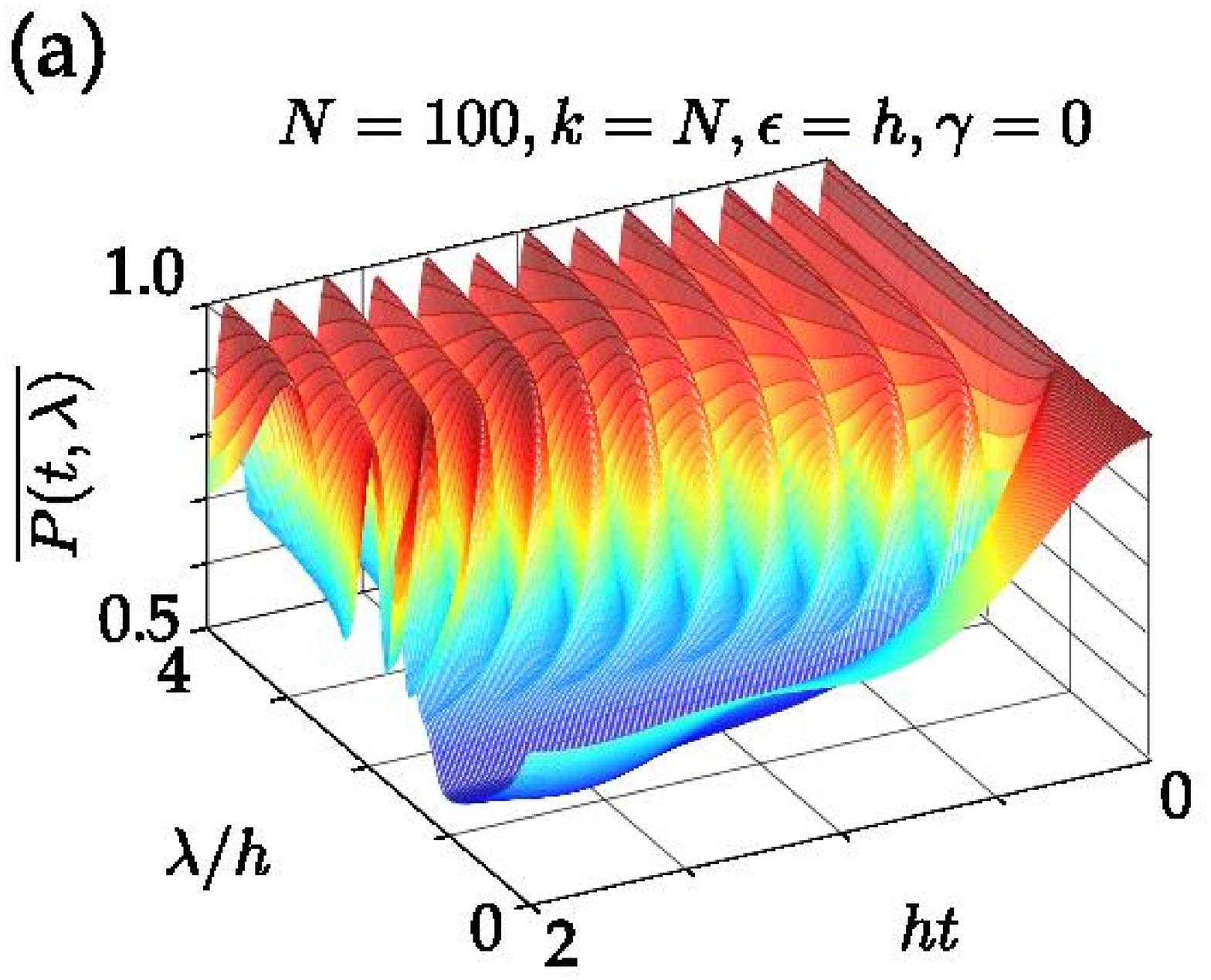}\hspace{0.25cm}
\includegraphics[width=4.1cm]{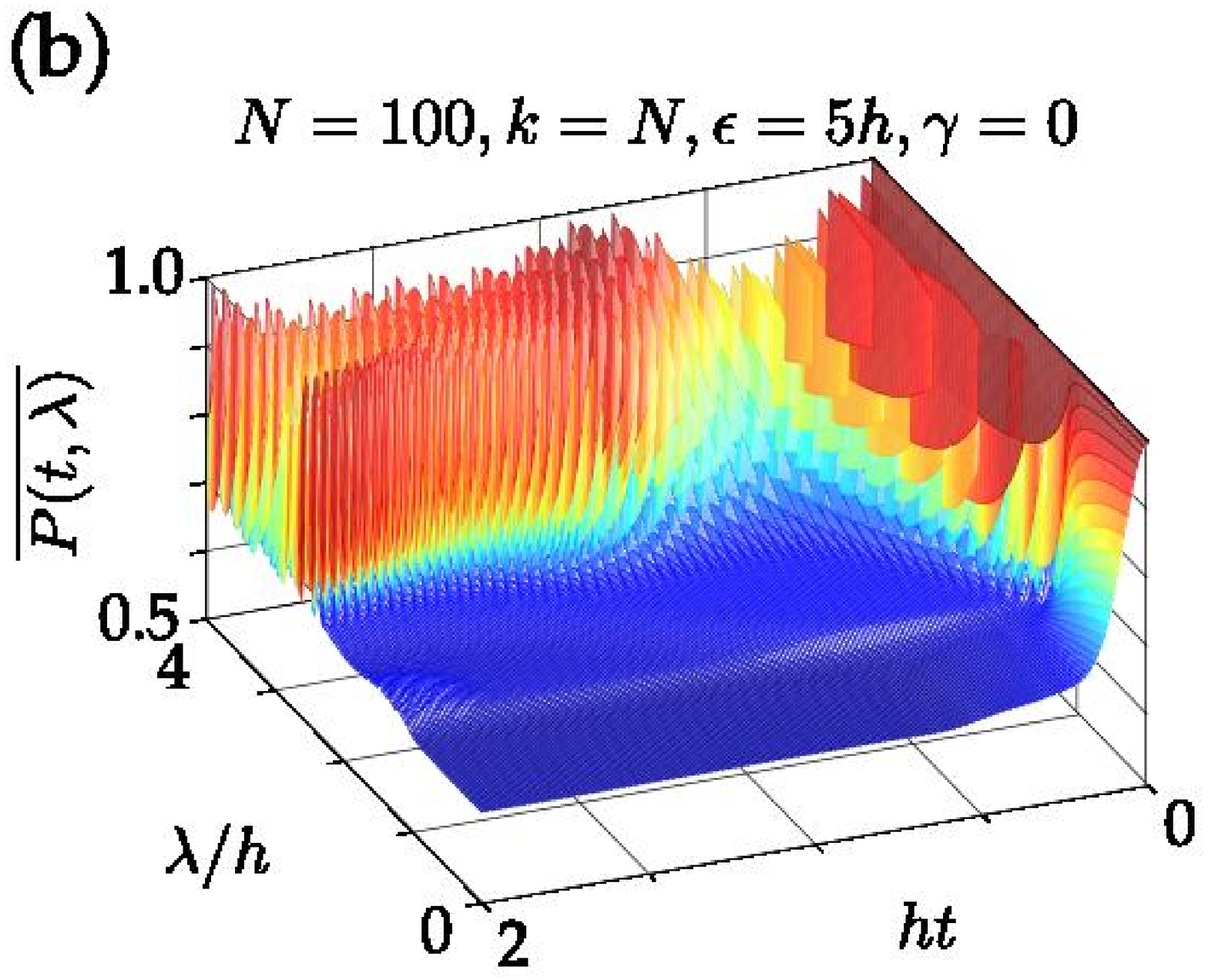}\vspace{0.3cm}
\includegraphics[width=4.1cm]{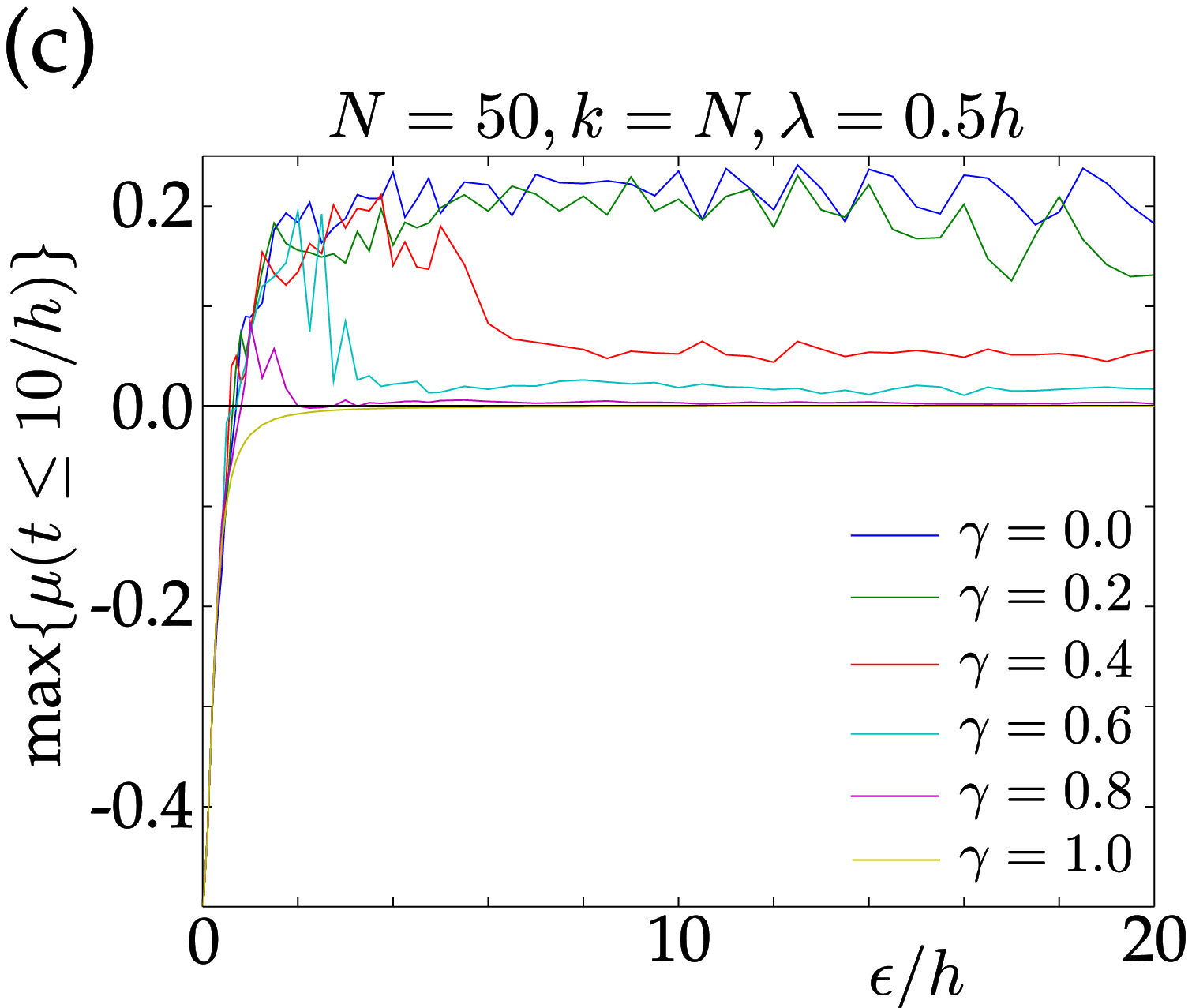}\hspace{0.25cm}
\includegraphics[width=4.1cm]{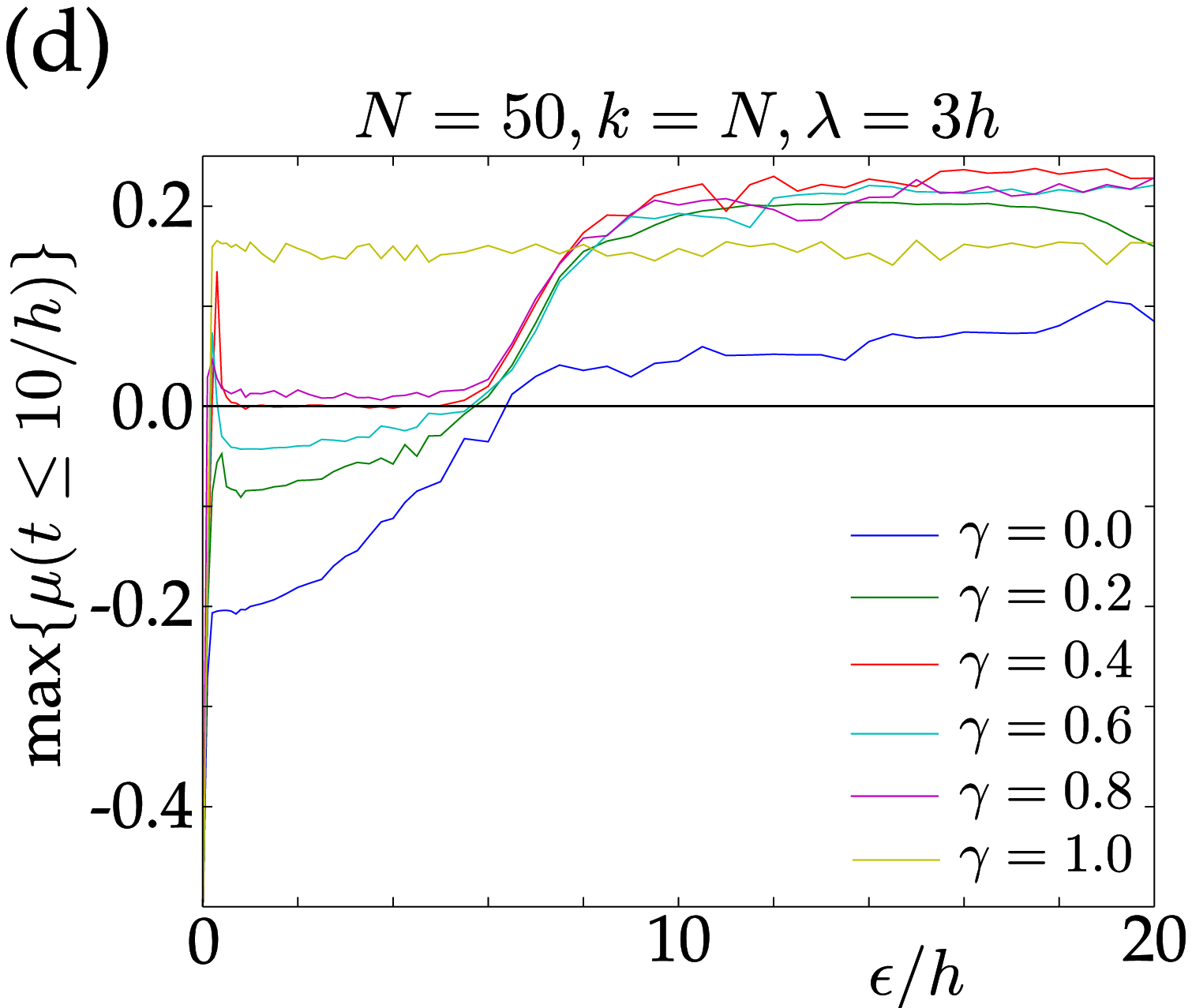}
\caption{LMG interaction, completely connected ($k=N$): (a) and (b),
the average purity plotted against $\lambda$ and $t$ for $\gamma=0$
and interaction strengths $\epsilon=h$ and $\epsilon=5h$
respectively. (c) and (d), the maximum value of $\mu (t)$ within
$t\leq 10/h$ at $\lambda=0.5h$ and $\lambda=3h$ respectively, as a
function of $\epsilon$ and for various values of $\gamma$. Values
above zero indicate that the evolution is entanglement breaking.}
\label{lmg_N}
\end{figure}

Figures \ref{lmg_N}(a) and \ref{lmg_N}(b) show the average purity,
as a function of intra-bath coupling strength and time, for a
completely connected qubit with $\gamma=0$ and interaction strengths
$\epsilon=h$ and $\epsilon=5h$ respectively. For this bath anistropy
there is a decay of the average purity in the normal phase even at
low interaction strengths. Note that the decay also occurs at
$\lambda=0$ for which the bath spins do not interact with each
other, and extends across criticality into the broken phase at high
interaction strengths $\epsilon\gtrsim h$. Rephasing only occurs on
short-time scales in the limit $\gamma\rightarrow 1$, with the
rephasing time $\tau_r$ independent of bath size for large baths,
$N\gtrsim100$. The interaction is now entanglement breaking in this
phase on short-time scales except in the limit $\gamma\rightarrow1$,
as shown in Fig. \ref{lmg_N}(c).

Approximate revivals occur in the broken phase away from criticality
with a frequency much greater than that of the corresponding single
link scenarios. Contrary to the Ising case, the rephasing time
$\tau_r$ increases with interaction strength and converges less
rapidly towards a finite value with increasing bath size. Therefore,
the rephasing time can only be considered independent of bath size
for very large baths, $N\gtrsim1000$. Figure \ref{lmg_N}(d) shows
that the induced decoherence is now periodically entanglement
breaking on short-time scales for all $\gamma$ above
$\epsilon\sim6h$ in this phase. However, at higher interaction
strengths this is no longer the case; for $\gamma=0$ the decoherence
is not entanglement breaking for $\epsilon>23h$ (see Fig.
\ref{lmg_k}(d)).

\subsection{Multiple links}

\begin{figure}
\includegraphics[width=4.1cm]{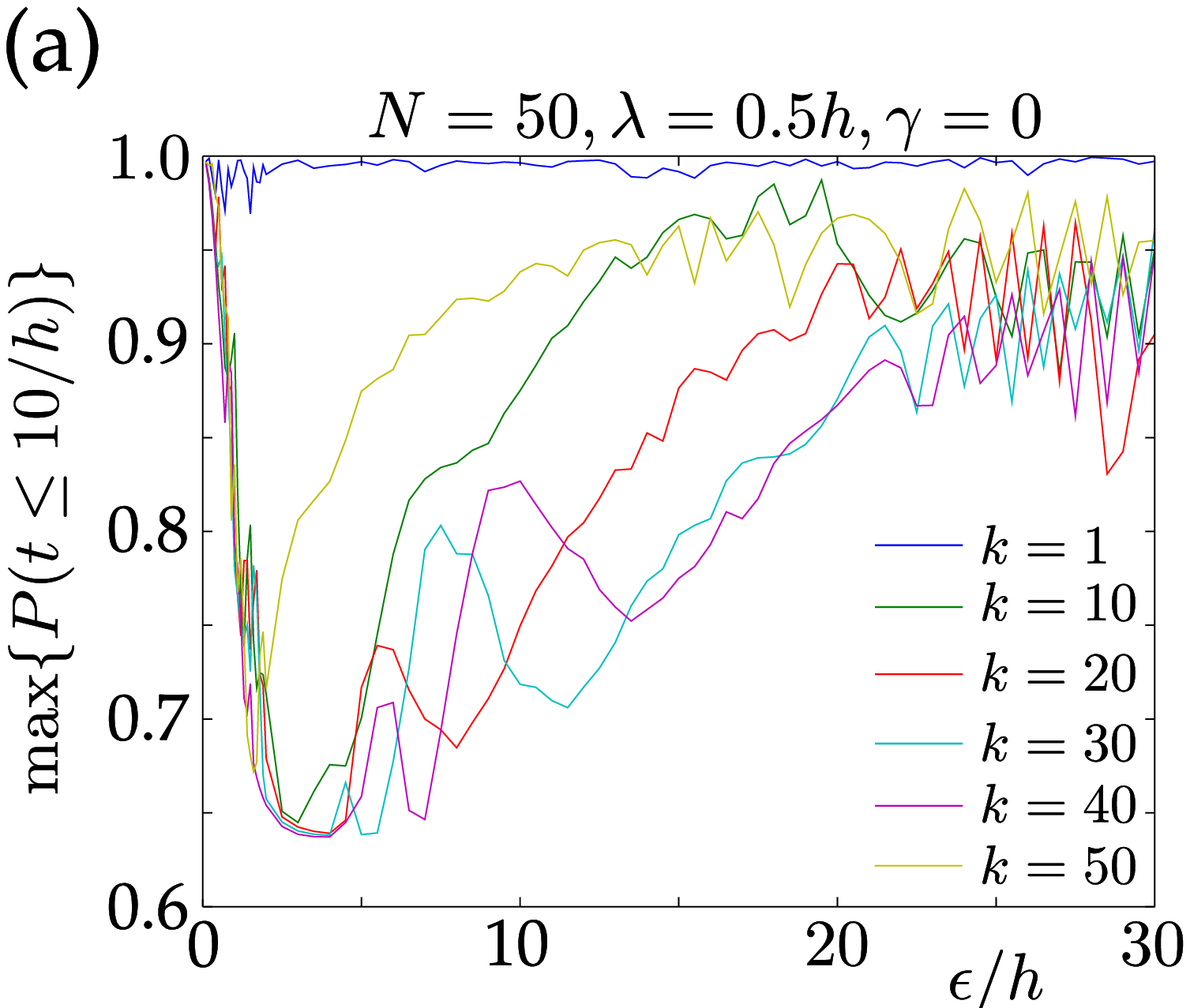}\hspace{0.25cm}
\includegraphics[width=4.1cm]{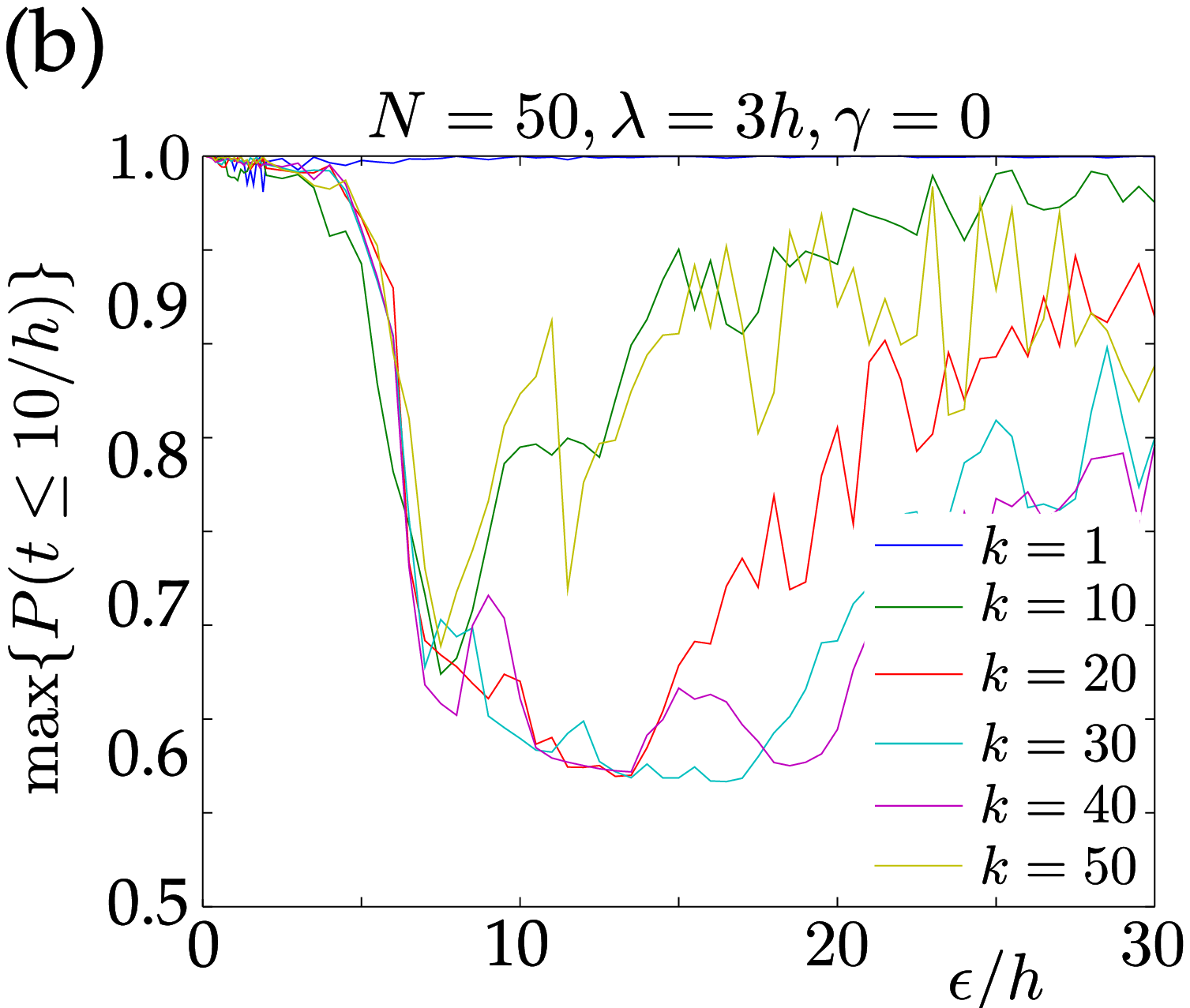}\vspace{0.3cm}
\includegraphics[width=4.1cm]{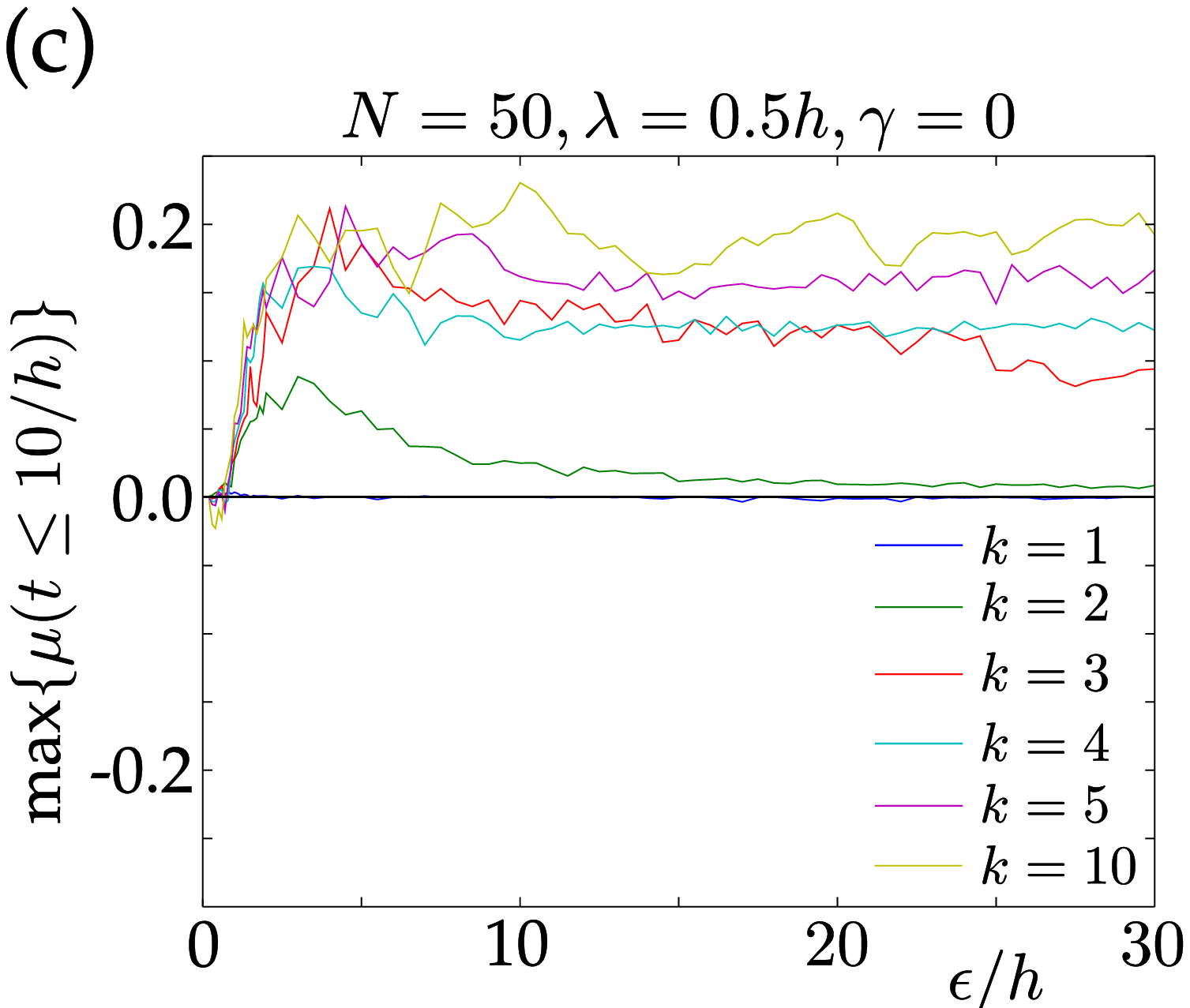}\hspace{0.25cm}
\includegraphics[width=4.1cm]{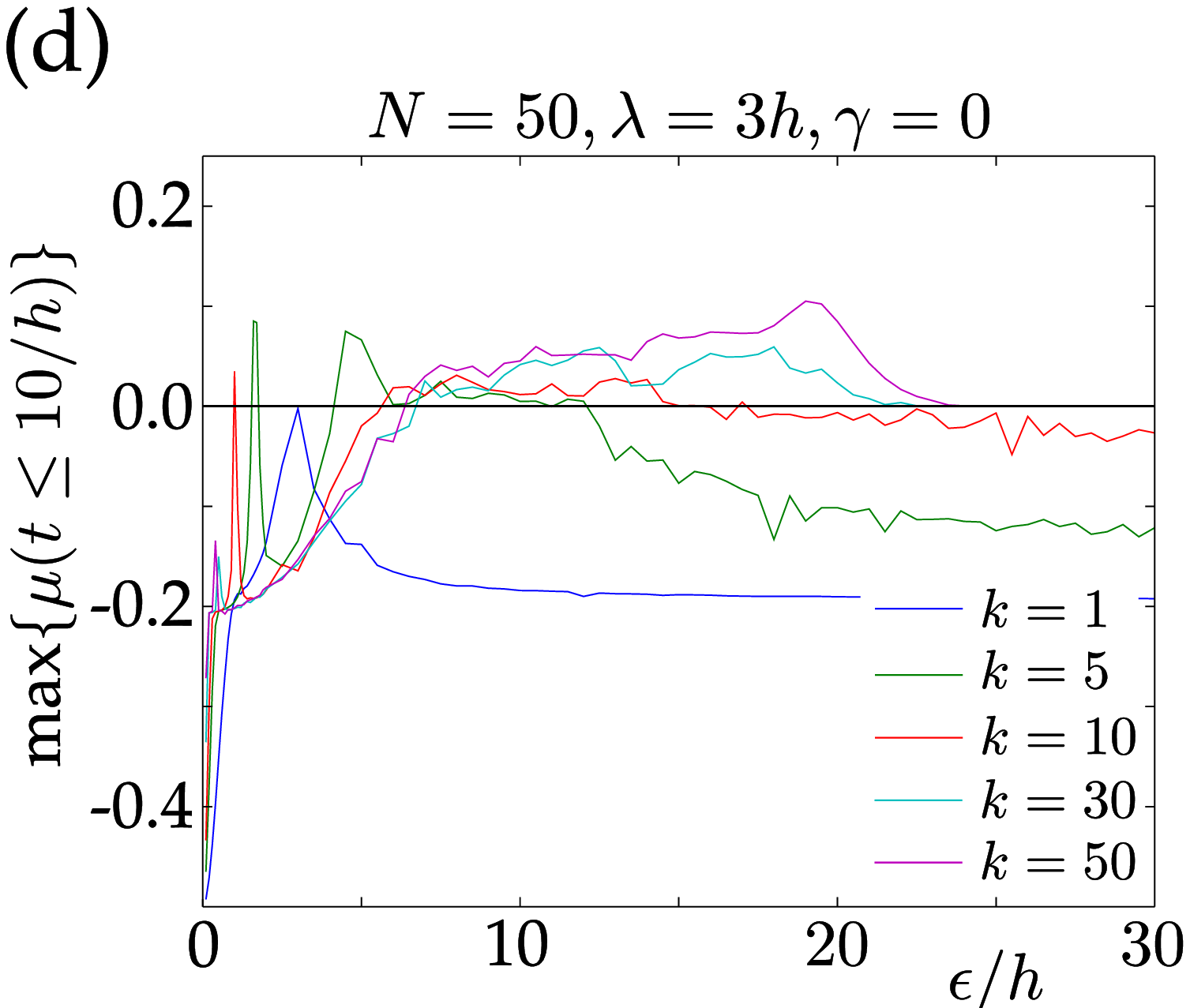}
\caption{LMG interaction: (a) and (b), the maximum value of the
purity at $\lambda=0.5h$ and $\lambda=3h$ respectively, for
rephasing within a time $t\leq 10/h$, as a function of interaction
strength $\epsilon$ and for various values of $k$. (c) and (d), the
maximum value of $\mu (t)$ within $t\leq 10/h$ at $\lambda=0.5h$ and
$\lambda=3h$ respectively, as a function of $\epsilon$ and for
various values of $k$. Values above zero indicate that the evolution
is entanglement breaking.} \label{lmg_k}
\end{figure}

As was the case for the Ising interaction, for the dissipative LMG
interaction we observe a change in the qubit's behavior as we move from
a single link to fully connected to the bath. In this subsection we
discuss how this transition occurs as we vary the link number $k$
for $\gamma=0$. For this bath anisotropy approximate rephasing
occurs on short-time scales in both phases for a single link, but is
suppressed in the normal phase for $k=N$. Also, the decoherence is
never entanglement breaking for $k=1$ with $\gamma=0$, but is at
high interaction strengths for $k=N$. We consider intra-bath
coupling strengths away from criticality of $\lambda=0.5h$ for the
normal phase and $\lambda=3h$ for the broken phase.

Figures \ref{lmg_k}(a) and \ref{lmg_k}(b) show the maximum value of
the average purity for rephasing within $t\leq10/h$ with a bath of
$N=50$ spins. Both figures are qualitatively similar to Fig.
\ref{Ising_k}(a) for the Ising interaction; approximate rephasing
occurs for a single link at any interaction strength, whilst the
maxima in the average purity are suppressed at high interaction
strengths for greater link numbers. In the broken phase, high
average purities of $\overline{P}>0.98$ are achieved for any link
number below $\epsilon\sim4h$, whilst in the normal phase this isn't
the case and there is an immediate decay of the average purity above
$\epsilon=0$ for $k\neq1$. As for the Ising interaction, the
suppression of maxima is reduced as the interaction strength is
further increased, with the rate of increase slowest for link
numbers $k\sim30$-$40$. This suggests that our analysis discussed in
Sec. \ref{Ising_multiple}, where the induced decoherence was limited
by the size of the subspace required to describe the partitioned
bath states, has a broader applicability than the Ising interaction
alone. Rephasing depends on the number of significant Fourier
components for, in this case, the average purity. This quantity is
reduced in the limits $\epsilon\ll h$ and $\epsilon\gg h$ for which
the interaction term is a small perturbation on the bath Hamiltonian
and vice versa respectively.

In Figs. \ref{lmg_k}(c) and \ref{lmg_k}(d) we have plotted the
maximum value of $\mu (t)$ within $t\leq 10/h$ for the normal and
broken phases respectively. In the normal phase entanglement
breaking occurs for $k\neq1$ at interaction strengths
$\epsilon>0.8h$, whilst in the broken phase the induced decoherence
is entanglement breaking for $k\neq1$ within the range
$h<\epsilon<23h$. It is only the case of coupling via a single link
that the qubit is immune to entanglement breaking within the period
$t\leq10/h$.

\section{Conclusions}
\label{conclusion}

To summarize, we have investigated the quantum evolution of a single
qubit coupled to a Lipkin-Meshkov-Glick bath as a model for
decoherence in solid state quantum memories. The bath, which
exhibited a second-order quantum phase transition, was highly
symmetric and allowed for exact calculations of the system dynamics
for large system sizes, $N\sim100$ spins. Further, partitioning of
the bath was possible without significantly increasing the
computational complexity, which allowed us to determine the effect
of increasing exposure of the qubit to the bath spins. Decoherence
of the qubit was quantified using the average purity and determining
if and when the evolution destroys any entanglement the qubit may
have with an external subsystem.

For the qubit interacting via an Ising interaction with just a few
bath spins, we observed zero dephasing in the normal phase and
almost complete revivals of qubit coherence in the broken phase. The
rephasing time, which was sensitive to criticality in the bath, was
independent of bath size for large baths and thus faithful qubit
storage would be possible in the thermodynamical limit. In general,
as the number of links was increased the revivals in the purity were
suppressed, which was discussed in terms of the number of components
in the Fourier Transform of the decoherence factor. Decoherence was
suppressed by the restrictive size of the subspace required to
describe the partitioned bath states, which was smallest for just a
few links. Further, it is likely that revivals were observed in
certain scenarios, even in the thermodynamical limit, because the
symmetry of the bath constrained the dynamics to a subspace whose
dimension grows only linearly with system size. Such revivals could
possibly occur in other systems possessing exchange symmetry that
are not as easily amenable to analysis as the LMG model. However, they may be difficult to observe experimentally due to a potential lack of the required bath symmetry in real samples.

In contrast to the Ising interaction, for the dissipative XY-type
(LMG) interaction between the qubit and the bath, decoherence was
dependent on the energy difference between the two levels of the
qubit and additionally was observed in both bath phases. Revivals
were found to occur for certain parameter regimes and just a few
links between the qubit and the bath. These were once again
suppressed as the link number was increased, broadening our
analytical discussion for the case of the Ising interaction in terms
of the restrictive size of the bath's Hilbert space. Interestingly,
the revivals occurred after periods when the evolution was
entanglement breaking, indicating a constant transfer of quantum
information back and forth between the qubit and bath. Our results
were based on a worst-case regime in which the qubit was subjected
to an identical transverse field to the bath and in general was
found to be less robust against decoherence than the Ising
interaction. Thus, systems with Ising interactions coupling only to
a small neighborhood of the environment would perform better as
quantum memories.

\begin{acknowledgments}
This work was supported by the UK EPSRC through projects QIPIRC
(GR/S82176/01) and EuroQUAM (EP/E041612/1).
\end{acknowledgments}

\appendix

\section{Quantum Operations Formalism and the Jamiolkowski Isomorphism}
\label{QOFandJI}

A powerful tool for considering the evolution of a quantum system is
the quantum operations formalism \cite{NielsenChuang}. If the
initial state of the quantum system is described in terms of a
density operator $\rho$, the subsequent evolution causes a
transformation to a final state given by the mapping
$\rho\rightarrow\varepsilon\{\rho\}$. The linear, completely
positive map $\varepsilon$ is known as a quantum operation. For a
closed system that implements a particular unitary operation $U$,
the quantum operation is simply $\varepsilon(\rho)=U\rho U^\dagger$.
In the same context, the evolution of an open quantum subsystem
interacting with some environment $R$, e.g. a single spin-1/2
interacting with an LMG spin bath, can be described as
\begin{equation}
\varepsilon(\rho)=\textrm{Tr}_R[U_T(\rho\otimes\rho_R)U_T^\dagger]
\textrm{ ,} \label{quantumoperation}
\end{equation}
where we consider the total system as closed and obtain the final
reduced density operator for the principal subsystem by tracing out
the state of the environment (note that we assume an initial product
state for the subsystem and environment).

Although an elegant description, calculating the final state in the
above way is often computationally difficult due to the large size
of the total Hilbert space. We can instead express Eq.
(\ref{quantumoperation}) explicitly in terms of the principal
subsystem's Hilbert space $\mathcal{H}$ by
\begin{equation}
\varepsilon(\rho) = \sum_{i=1}^{d^2}{A_i\rho A_i^\dagger} \textrm{
.} \label{kraus}
\end{equation}
where the $A_i$, which act on $\mathcal{H}$, are known as Kraus
operators and $d$ is the dimension of $\mathcal{H}$. Once the Kraus
operators are known for a particular type of evolution, one can
easily obtain the final state of the open subsystem given any
initial state. Importantly, the Kraus operators reveal the nature of
the noise induced by the coupling to the environment. For a unital
map, which is always the case for the noise described by the
coupling to an environment, the Kraus operators satisfy
$\sum_{i}{A_iA_i^\dagger}=\openone$.

An equivalent description of the evolution in Eq. (\ref{kraus}) can
be made in terms of a superoperator $\Lambda$. To explain this more
fully, consider the Hilbert space $\mathcal{H}$ spanned by basis
states $\{|i\rangle\ |\ i=0,\cdots ,d-1\}$. We can expand any
density operator $\rho$ for the system in the operator basis
$\{|i\rangle\langle j|\ |\ i,j=0,\cdots ,d-1\}$, with its
corresponding matrix elements $\rho_{ij}$ contained in a
$d^2$-dimensional vector. The superoperator $\Lambda$ can then be
described by a $d^2\times d^2$ super-matrix with elements
\begin{equation}
\Lambda = \sum_{kl=0}^{d-1}{\Lambda_{ij,kl}|k\rangle\langle l|}
\textrm{ .} \label{lambda}
\end{equation}
Note that $\Lambda$ can be inferred from the Kraus operators and
vice versa via
\begin{equation}
\Lambda = \sum_{i=1}^{d^2}{A_i^*\otimes A_i} \textrm{ .}
\label{lambdakraus}
\end{equation}

The Jamiolkowski isomorphism \cite{Jamiolkowski.3.275} exploits an
initial setup of two copies of the principal subsystem, such that
the subsequent evolution transfers all the information about
$\Lambda$ to the final quantum state. The two copies $a$ and $b$ are
prepared in the maximally entangled state $|\Psi^+\rangle
=\tfrac{1}{\sqrt{d}}\sum_{i=0}^{d-1}{|i\rangle\otimes |i\rangle}$ and the superoperator $\Lambda$ is applied to $b$ as
\begin{equation}
(\openone\otimes\Lambda)\{|\Psi^+\rangle\langle\Psi^+|\} =
\frac{1}{d}\sum_{ij=0}^{d-1}{|i\rangle\langle
j|\otimes\Lambda|i\rangle\langle j|} = \rho^\Lambda \textrm{ .}
\label{jamiolkowski}
\end{equation}
The matrix elements of the resulting density operator $\rho^\Lambda$
are related to those of $\Lambda$ by $d\rho_{ikjl}^\Lambda =
\Lambda_{ijkl}$, where
\begin{equation}
\rho^\Lambda = \sum_{ijkl=0}^{d-1}{\rho_{ikjl}^\Lambda
|i\rangle\langle j|\otimes |k\rangle\langle l|} \textrm{ .}
\label{rho_lambda}
\end{equation}
Thus, using the Jamiolkowski isomorphism, it is sufficient to
calculate $\rho^\Lambda$ to determine the superoperator $\Lambda$,
and then trivially the Kraus operators $A_i$, for any quantum
operation $\varepsilon(\rho)$.

\section{The Loschmidt Echo for a single link and an isotropic bath}
\label{appendix_k=1}

The GS of an isotropic, ferromagnetically coupled bath is the Dicke
state $|N/2,M\rangle$, where $M=N/2\textrm{, }\lfloor
hN/2\lambda\rceil$ in the normal and broken phases respectively. For
$N\gg 1$ the latter can be approximated as $M=hN/2\lambda$. When we
partition the bath into a single and $N-1$ spins the Dicke state
$|N/2,M\rangle$ decomposes into two terms
\begin{equation}
\begin{split}
\left|\tfrac{N}{2},M\right\rangle = &\ c_{1/2}\left|\tfrac{1}{2},\tfrac{1}{2}\right\rangle\otimes\left|\tfrac{N}{2}-\tfrac{1}{2},M-\tfrac{1}{2}\right\rangle \\
&
+c_{-1/2}\left|\tfrac{1}{2},-\tfrac{1}{2}\right\rangle\otimes\left|\tfrac{N}{2}-\tfrac{1}{2},M+\tfrac{1}{2}\right\rangle
\textrm{ ,}
\end{split}\label{k=1 decomp}
\end{equation}
where the Clebsch-Gordon coefficients are given by
\begin{equation}
\begin{split}
c_{1/2} = &\ \left\langle\tfrac{1}{2},\tfrac{1}{2}\right|\otimes\left\langle\tfrac{N}{2}-\tfrac{1}{2},M-\tfrac{1}{2}\middle|\tfrac{N}{2},M\right\rangle = \sqrt{\tfrac{1}{2}+\tfrac{M}{N}} \textrm{ ,} \\
c_{-1/2} = &\
\left\langle\tfrac{1}{2},-\tfrac{1}{2}\right|\otimes\left\langle\tfrac{N}{2}-\tfrac{1}{2},M+\tfrac{1}{2}\middle|\tfrac{N}{2},M\right\rangle
= \sqrt{\tfrac{1}{2}-\tfrac{M}{N}} \textrm{ .}
\end{split}
\end{equation}
Evolution of the GS under the perturbed bath Hamiltonian
$H_-=H_R-\epsilon\sigma_1^z$ for this scenario is restricted to the
two states on the RHS of Eq. (\ref{k=1 decomp}). Using these states
as a basis, the action of $H_-$ on the Dicke state $|N/2,M\rangle$
can be summarised by the matrix equation
\begin{equation}
H_-\left|\tfrac{N}{2},M\right\rangle = h
\begin{pmatrix}
\alpha-\epsilon/h & \beta \\
\beta & \alpha+\zeta+\epsilon/h
\end{pmatrix}
\begin{pmatrix}
c_{1/2} \\
c_{-1/2}
\end{pmatrix} \textrm{ ,} \label{H_- action}
\end{equation}
where
\begin{gather}
\alpha = -\frac{\lambda}{2hN}\left\{(N-1)^2-(2M-1)^2\right\} - 2M \textrm{ ,} \\
\beta = -\frac{\lambda}{h}\sqrt{1-\frac{4M^2}{N^2}} \textrm{ ,} \\
\zeta = \frac{4\lambda M}{hN} \textrm{ .}
\end{gather}
For $N\gg 1$, the latter two quantities become
\begin{gather}
\beta = -\frac{\lambda}{h}\sqrt{1-\frac{h^2}{\lambda^2}} \textrm{ ,} \\
\zeta = 2 \textrm{ .}
\end{gather}
By firstly diagonalizing the matrix representation of $H_-$ in Eq.
(\ref{H_- action}), we can find a similar representation for
$e^{-iH_-t}$ in this basis, given by
\begin{widetext}
\begin{equation}
e^{-iH_-t} = \frac{e^{-i(\alpha+1)ht}}{\eta_-}
\begin{pmatrix}
\eta_-\cos{(\eta_- t/2)} + 2i(h+\epsilon)\sin{(\eta_- t/2)} & -2i\beta h\sin{(\eta_- t/2)} \\
-2i\beta h\sin{(\eta_- t/2)} & \eta_-\cos{(\eta_- t/2)} -
2i(h+\epsilon)\sin{(\eta_- t/2)}
\end{pmatrix} \textrm{ ,}
\label{eH_- action}
\end{equation}
where $\eta_-=2\sqrt{\lambda^2+\epsilon(\epsilon+2h)}$. Similarly,
for $e^{-iH_+t}$ we obtain
\begin{equation}
e^{-iH_+t} = \frac{e^{-i(\alpha+1)ht}}{\eta_+}
\begin{pmatrix}
\eta_+\cos{(\eta_+ t/2)} + 2i(h-\epsilon)\sin{(\eta_+ t/2)} & -2i\beta h\sin{(\eta_+ t/2)} \\
-2i\beta h\sin{(\eta_+ t/2)} & \eta_+\cos{(\eta_+ t/2)} -
2i(h-\epsilon)\sin{(\eta_+ t/2)}
\end{pmatrix} \textrm{ ,}
\label{eH_+ action}
\end{equation}
\end{widetext}
where $\eta_+=2\sqrt{\lambda^2+\epsilon(\epsilon-2h)}$. The LE for
the broken phase ($\lambda>h$) can now be calculated by multiplying
the hermitian conjugate of Eq. (\ref{eH_+ action}) by Eq. (\ref{eH_-
action}) and taking the expectation value with respect to the GS,
i.e.
\begin{multline}
L(t)=\Bigg| \left(c_{1/2},c_{-1/2}\right) e^{iH_+t}e^{-iH_-t}
\begin{pmatrix}
c_{1/2} \\
c_{-1/2}
\end{pmatrix} \Bigg|^2
\textrm{ .} \label{LE1}
\end{multline}
The resulting expression for the LE is complicated and need not be
written here explicitly. Importantly, we can see revivals of full
qubit coherence occur at times $t=l\tau_c$ that satisfy both
$\eta_-\tau_c=2l_-\pi$ and $\eta_+\tau_c=2l_+\pi$, where $l$, $l_-$
and $l_+$ are all integers; at these times the matrices in Eq.
(\ref{eH_- action}) and Eq. (\ref{eH_+ action}) are both equal to
the identity. In the limit $\epsilon\ll h$ the two frequencies are
approximately equal and the coherence time is $\tau_c=\pi/\lambda$.

\bibliography{LMG-Paper}

\end{document}